\documentclass[11pt, a4paper]{article}
\pdfoutput=1
\usepackage{jcappub}
\usepackage{natbib}
\usepackage{aas_macros}

\usepackage[T1]{fontenc}
\usepackage[utf8]{inputenc}
\usepackage[english]{babel}
\usepackage{amsmath}
\usepackage{amssymb}

\usepackage{graphicx}
\usepackage{epsfig,epstopdf}
\usepackage{color,xcolor}
\usepackage{multirow}
\usepackage{mleftright}

\usepackage{url}
\usepackage{xspace}
\usepackage{comment}
\usepackage{bm}
\usepackage{braket}
\usepackage{cleveref}
\usepackage{hhline}

\usepackage{hyperref}
\newcommand{\bk}{\mathbf{k}}
\newcommand{\bq}{\mathbf{q}}
\newcommand{\bx}{\mathbf{x}}
\newcommand{\fnl}{f_{\rm NL}}
\newcommand{\fnll}{f_{\mathrm{NL}}^{\mathrm{loc}}}
\newcommand{\fnle}{f_{\mathrm{NL}}^{\mathrm{equil}}}
\newcommand{\fnlo}{f_{\mathrm{NL}}^{\mathrm{orth}}}
\newcommand{\hoMpc}{\ensuremath{\text{$h$/Mpc}}\xspace}
\newcommand{\kpar}{k_{\parallel}}
\newcommand{\kparmin}{k_{\parallel, {\rm min}}}
\newcommand{\kperp}{k_{\bot}}
\newcommand*\mean[1]{\overline{#1}}

\newcommand*{\order} [1] {\ensuremath{\mathcal{O}(#1)}\xspace}
\newcommand*{\ie} {i.\,\!e.\xspace}
\newcommand*{\eg} {e.\,\!g.\xspace}

\newcommand*{\veps} {\varepsilon}
\newcommand*{\eref} [1] {Eq.\ (\ref{#1})}

\newcommand*{\tref} [1] {Table \ref{#1}\xspace}
\newcommand*{\sref} [1] {Sec.\ \ref{#1}\xspace}
\newcommand*{\fref} [1] {Fig. \ref{#1}\xspace}
\newcommand*{\aref} [1] {Appendix \ \ref{#1}\xspace}
\newcommand*{\perm} {\;\text{perm}}

\newcommand{\Puma}{\text{PUMA}}

\graphicspath{{./plots/}}

\title{Forecasts on Primordial non-Gaussianity from 21\,cm Intensity Mapping experiments}

\author[a,c,1]{Dionysios Karagiannis,\note{Corresponding author.}}
\author[b]{An\v{z}e Slosar,}
\author[a,c]{Michele Liguori,}

\affiliation[a]{Dipartimento di Fisica e Astronomia “G. Galilei”,Università degli Studi di Padova, via Marzolo 8, I-35131, Padova, Italy}
\affiliation[b]{Physics Department, Brookhaven National Laboratory, Upton, NY 11973, USA}
\affiliation[c]{INFN, Sezione di Padova, via Marzolo 8, I-35131, Padova, Italy}

\emailAdd{dakaragian@gmail.com}

\abstract{
  We forecast the ability of dedicated 21\,cm intensity mapping experiments to constraint primordial non-Gaussianity using power spectrum and bispectrum. We model the signal by including the non-linear biasing expansion through a generalized halo model approach. We consider the importance of foreground filtering scale and of the foreground wedge. We find that the current generation intensity mapping experiments like CHIME do not posses sufficient sensitivity to be competitive with the existing limits. On the other hand, upcoming experiments like HIRAX can improve the current constraints and the proposed PUMA experiment can substantially improve them, reaching sensitivities below $\sigma(\fnl)<5$ for equilateral and orthogonal configurations and $\sigma(\fnl) < 1$ for the local shape, if good foreground control is achieved. }

\begin{document}
% Abstract of the paper
 \maketitle
\flushbottom

\section{Introduction}\label{sec:INTRO}

Understanding primordial cosmological inflation in detail is one of the biggest challenges of modern physics. Currently, inflation is essentially a generic framework for setting up the hot big bang, rather than a concrete, specific theory:  any scalar field rolling down a sufficiently flat potential will produce a nearly spatially flat universe with small seed fluctuations which allow the hot Big Bang to proceed and eventually create the universe as we know it. Nevertheless, we do have a few observational handles that help us to distinguish between a simplest realization of inflation -- that is, a minimal single-field scenario -- and  more complex models, such as those involving multiple fields and non-minimal couplings. A convincing detection in any of these probes of non canonical physics will be crucial in connecting inflation to the bigger puzzle of fundamental physics.

Primordial non-Gaussianity (PNG) has emerged as one such powerful probe in the last decade \cite{2019BAAS...51c.107M, 1412.4671}. In the simplest models of inflation, containing a single scalar field, the resulting fluctuations in the curvature field are nearly perfectly Gaussian, essentially encoding the physics of the lowest energy state of a quantized scalar field. This picture breaks, if there are either multiple interacting fields involved in inflation, or non-minimal couplings of the inflaton, or if slow-roll is broken \cite{Bartolo2004b,Sasaki2006,Byrnes2008,Byrnes2009a,Byrnes2009b,Byrnes2010,Linde1996,Zaldarriaga2003,Creminelli2003,Seery2005} (see also Refs. \cite{Bartolo2004,Komatsu2009,1412.4671} for a general review and discussion). Since we know that the primordial fluctuations \emph{are} Gaussian to a very good degree, these effects must be perturbative. As a result, in most scenarios, a non-zero three-point function is generated at leading order in a cumulant expansion of the primordial curvature fluctuation field. Resulting bi-spectra are classified by the shape of the triangle configurations that dominate the signal, as we will discuss in more detail in the next section. Needless to say, a detection of this signal would bear far-reaching theoretical implications and be of paramount importance in cosmology.

Measurements of the very small non-Gaussianity signal are
difficult and dominated by the sample variance of the observable,
i.e. by the number of modes which can be used to measure any deviation
from zero bi-spectrum. Currently, the most competitive constraints
come from CMB observations \cite{1905.05697}. However, the constraints
derived from the Large Scale Structure (LSS) are catching up fast
\cite{Slosar2008,1208.1491,1303.1349,1405.4315,1310.6716,1904.08859}.
In fact, due to its inherently three-dimensional structure and thus a
larger number of measured linear modes, LSS is expected to soon
supersede CMB measurements \cite{Karagiannis2018,1904.08859}.

In light of this premise, we investigate here the possibility of 21\,cm intensity mapping (IM) observations to probe primordial non-Gaussianity. More specifically, we develop the methodology to perform non-Gaussianity forecasts for intensity mapping surveys, and apply it to current and proposed experiments. Developments in technology and in our understanding of the signal have recently allowed proposing very ambitious experimental designs in this area, which can still be implemented at reasonable cost. For example, PUMA \cite{PUMA_surv} will observe the low-redshift universe from $z=0.3$ to $z=6$, with a single instrument over approximately half the sky. This extremely large volume suggests the possibility that intensity mapping will be at least competitive with the CMB in placing primordial non-Gaussianity constraints. 

The forthcoming Square Kilometer Array (SKA) \cite{SKA_redbook} neutral hydrogen survey is expected to probe the cosmological HI signal over $3/4$ of the sky, for a wide range of redshifts, allowing us to push below the Planck constraints on PNG \cite{Camera2015}. In addition, the SKA phase 1, with its large volume and wide frequency range, could in principle produce competitive constraints on the local PNG, where uncertainties of $\sigma(\fnll)\approx 1-4$ are considered to be achievable \cite{Camera2013,Camera2015,Alonso2015,SKA_redbook}. These approaches utilise the large scale power excess, induced by PNG, in the power spectrum of biased tracers \cite{Dalal2008,Slosar2008,Matarrese2008,Afshordi2008,Verde2009} and treat SKA1 as a set of single dishes \cite{Santos2015}, while consider an effective treatment of the foregrounds (see Refs.~\cite{Alonso2015,Cunnington:2020wdu} for a discussion). These constraints could be pushed even further by taking into advantage the multitracer technique \cite{Fonseca2015,Alonso:2015sfa}. In Ref. \cite{Xu2015}, they use the Tianlai cylinder array \cite{Chen2012} to present forecasts on PNG from both power spectrum and bispectrum, where they assume a rather simple model for the latter. Here we adopt the complete model, up to tree-level, for the redshift space HI power spectrum and bispectrum, while we take into account theoretical limitations and a variety of observational effects (see Ref. \cite{Karagiannis2018} for an extensive discussion). Adding to this, a careful treatment of foreground systematics is applied, providing realistic forecasts on the amplitude of PNG.

We note here that it might seem fanciful to make forecasts for the ability of 21\,cm
experiments to measure non-Gaussianity, when we do not even have
basic clean auto-power spectrum measurements from observations of
21\,cm intensity maps, let alone baryon acoustic oscillation
measurements. However, we note that the current observational issues
are entirely due to imperfect calibration and stability of the
instrument, rather than being of fundamental astrophysical nature. Therefore,
the approach we take is to forecast the statistical sensitivity to
non-Gaussianity in the presence of irreducible systematic issues such
as foreground contamination, assuming that purely instrumental issues
will be solved in time. In fact, the competitive figures that we
find should provide additional impetus for the R\&D that is required
to make the 21\,cm observations a success.

This paper is structured as follows. In Section~\ref{sec:modeling-signal} we
present our model for the large-scale fluctuations of the neutral
hydrogen density. In Section~\ref{sec:surveys} we discuss observational limitations and specifications of three experiments under consideration in this paper, while in Section~\ref{sec:fisher} we present the Fisher matrix formalism used to forecast the amplitude of PNG. We present our results in Section~\ref{sec:RESULTS}, followed by Conclusions (Section~\ref{sec:CONCL}).

\section{Modeling the signal}
\label{sec:modeling-signal}
\subsection{Matter power spectrum and bispectrum}

The two point statistics of the primordial curvature perturbation field in Fourier space, $\zeta(\bk)$, is defined as

\begin{equation}
  \langle \zeta(\bk)\zeta(\bk')\rangle=(2\pi)^3 \delta_D(\bk-\bk')P_\zeta(\bk).  
\end{equation}
These primordial perturbations, generated through inflation, are nearly perfectly Gaussian. Therefore, they can be characterized by the two-point correlation function. They are also directly related to the Bardeen gauge invariant primordial gravitational potential $\Phi(\bk)$ (during matter domination era, $\Phi(\bk)=3/5\zeta(\bk)$), which is in turn related, through the cosmological Poisson equation, to the linearly evolved dark matter density contrast $\delta_m^L(\bk,z)=M(k,z)\Phi(k)$. The linear matter power spectrum is thus defined, as 
 
 \begin{equation}\label{eq:powlin}
  P_m^L(k,z)=M^2(k,z)P_\Phi(k),
 \end{equation}

 \noindent where 
 
 \begin{equation}\label{eq:poisM}
  M(k,z)=\frac{2k^2c^2T(k)D(z)}{3\Omega_m H_0^2}.
 \end{equation}
 
 \noindent In the above equation, $D(z)$ is the linear growth factor, originating from the growing mode of the linear fluid equations, normalised to unity today (\ie $D(0)=1$). $T(k)$ is the matter transfer function normalized to unity at large scales ($k \rightarrow 0$) and $c$ is the speed of light. In this work, we use {\it Planck} 2015 best-fit parameters \cite{Planck2016_cosmopar} to define the fiducial cosmology used to derive the matter power spectrum, which is computed with the numerical Boltzmann code CAMB \cite{CAMB}.
 
 Various inflationary theories predict a small deviation from perfectly Gaussian initial conditions. This leads to non-zero high-order correlation functions of the curvature perturbation field. The largest of such correlators -- in Fourier space -- is in most cases the bispectrum, i.e.,  the three-point correlation function of Fourier modes, defined as:
 
 \begin{equation}
   \langle\zeta(\bk_1)\zeta(\bk_2)\zeta(\bk_3)\rangle=(2\pi)^3\delta_{D}(\bk_1+\bk_2+\bk_3)B_{\zeta}(\bk_1,\bk_2,\bk_3).
  \end{equation}
 The strength of the bispectrum signal is generally described by a dimensionless amplitude parameter, called $f_{\rm NL}$.
 \noindent In addition to this, the Dirac delta (enforcing homogeneity) imposes a dependence of the bispectrum on the shape of the different Fourier space triangles. Following the equations above, we can write, at leading order, the linearly extrapolated PNG contribution of the matter density bispectrum as:

  \begin{equation} \label{eq:bisng}
   B_I(k_1,k_2,k_3,z)=M(k_1,z)M(k_2,z)M(k_3,z)B_\Phi(k_1,k_2,k_3) \,
  \end{equation}
  
    \noindent where $B_\Phi$, as in the case of the power spectrum, is related to $B_\zeta$ through the Poisson equation and therefore provides a window to the non-linear interaction during inflation. The number of shapes of the forming triangles is large and the different inflation models predict PNG that picks at different configurations. In this work, we will consider three very important shapes of PNG, namely the {\em local} shape \cite{Salopek1990,Gangui1993,Verde1999,Komatsu2001}, the {\em equilateral shape} \cite{Creminelli2005} and the {\em orthogonal} \cite{Senatore2009}, defined respectively as:
    
    \begin{align}
B_{\Phi}^{\text{loc}}(k_1,k_2,k_3)&=2 \fnll\left[P_{\Phi}(k_1)P_{\Phi}(k_2)+\text{2 perms} \right] \, ,\\
B_{\Phi}^{\text{eq}}(k_1,k_2,k_3)&=6 \fnle\big[-[P_{\Phi}(k_1)P_{\Phi}(k_2)+\text{2 perms} ] -2[P_{\Phi}(k_1)P_{\Phi}(k_2)P_{\Phi}(k_3)]^{2/3} \nonumber \\
&+[P_{\Phi}^{1/3}(k_1)P_{\Phi}^{2/3}(k_2)P_{\Phi}(k_3)+\text{5 perms}] \big] \,, \\ 
  B_\Phi^\text{orth}(k_1,k_2,k_3) &= 6\fnlo\big[ 3[P_\Phi^{1/3}(k_1)P_\Phi^{2/3}(k_2)P_\Phi(k_3) +5\text{ perms}] \nonumber \\
  &-3 \left[P_{\Phi}(k_1)P_{\Phi}(k_2)+\text{2 perms} \right]-8 (P_\Phi(k_1)P_\Phi(k_2)P_\Phi(k_3))^{2/3}
\big] \,.
\end{align}
The signal of these templates peaks at the squeezed triangles ($k_3\ll k_2\simeq k_1$), the equilateral configurations ($k_3\simeq k_2\simeq k_1$) and in both equilateral and folded triangles ($k_1\simeq k_2 \simeq k_3/2$), respectively.

Due to the non-linear nature of gravity, the matter bispectrum has additional terms at the zeroth-order (tree-level). Therefore, measuring the amplitude of PNG from the bispectrum of LSS is a highly non-trivial process. Robust modelling of non-linearities should be considered, in order to remove these gravitational contaminants and retrieve a clear PNG signal. Here, we account for the gravity induced non-linearities in the framework of Standard Perturbation Theory (SPT) [\eg see Ref. \cite{Bernardeau2002} for a review]. Through out this work, we will restrict our analysis up to linear/semi-nonlinear scales, in order to be consistent with the SPT approach. Thus, we will only use the linear power spectrum and the tree-level bispectrum
in SPT (see \sref{sec:power_bisp}).

\subsection{Halo bias and Mass function}\label{sec:BIAS}

The measurement of PNG from the LSS of the Universe requires a robust description of the relation between the dark matter haloes and the underlying matter distribution. The connection between the two is incorporated under the concept of \emph{bias}. A perturbative approach has been used in the past to describe the halo bias, by expanding the halo over-density field $\delta_h$ in powers of $\delta_m$ \cite{Coles1993,Fry1993,Fry1996,Catelan1997}, while tidal field terms were added in the expansion later on \cite{Catelan2000,McDonald2009,Elia2010,Chan2012,Baldauf2012}. A complete set of bias terms was derived in Refs.~\cite{Assassi2014,Senatore2014,Mirbabayi2014}, where $\delta_h$ is expanded over renormalised operators that incorporate all possible local gravitational observables, like $\delta_m$ or the tidal field $s_{ij}=(\partial_i\partial_j/\nabla^2-\delta_{ij}/3)\delta_m$. In this \emph{general bias} expansion and for Gaussian initial conditions, the terms up to second order (needed for the tree-level bispectrum) can be written in the Eulerian framework as  \cite{Assassi2014,Senatore2014,Mirbabayi2014}:

  \begin{equation}\label{eq:deltaG}
   \delta_h^{E,(G)}(\bx,\tau)= b_1^E(\tau)\delta(\bx,\tau) +\veps^E(\bx,\tau)+\frac{b_2^E(\tau)}{2}\delta^2(\bx,\tau) + \frac{b_{s^2}^E(\tau)}{2}s^2(\bx,\tau)+\veps_{\delta}^E(\bx,\tau)\delta(\bx,\tau) \,,
  \end{equation}
  where $\tau$ is the conformal time and $\bx$ are the spatial co-moving coordinates in the Eulerian frame. In addition, $s^2=s_{ij}s^{ij}$ is the simplest scalar that can be formed by the tidal field, $\veps^E$ is the leading stochastic bias contribution \cite{Dekel1998,Taruya1998,Matsubara1999} and $\veps_{\delta}^E$ is the stochastic counterpart of the linear bias. For the scales considered here, i.e. much larger scales than those involved in the halo formation, where the perturbative description breaks down, the terms in \eref{eq:deltaG} are sufficient to describe the halo statistics. This means that, the higher-order derivative contribution, which become important towards the halo formation scales, can be safely excluded (see Refs.~\cite{Desjacques2016,Desjacques:2018pfv} for the full modeling and extensive discussion). The second-order tidal field bias coefficient, following the convention of Ref.~ \cite{Baldauf2012}, is given by $b_{s^2}^E=-4/7(b_1^E-1)$. This relation assumes the Lagrangian tidal bias to be $b_{s^2}^L=0$, as well as a convolution of matter and tracer. It is tested against numerical results in Ref. \cite{Lazeyras2018}, where they find it to be a good approximation with an evidence of a small negative Lagrangian bias, \ie $b_{s^2}^L<0$.

  The presence of PNG introduces a scale-dependent bias correction to the linear bias, originating from the coupling between the long-wavelength fluctuations and the small scales, which in turn is induced by the presence of the primordial bispectrum in the squeezed limit. This effect was studied extensively for local PNG \cite{Dalal2008,Slosar2008,Matarrese2008,Afshordi2008,Verde2009,Desjacques2010}, whose template peaks in the squeezed triangle configurations. Taking the squeezed limit of the bispectrum, in the case of a general non-local quadratic non-Gaussianity (\eg equilateral PNG), also generates a scale-dependent bias correction \cite{Schmidt2010,Scoccimarro2011,Desjacques2011b,Schmidt2013}. In order to model this effect into the general bias description, an additional field $\Psi$ is added (in the same spirit as in Refs.~\cite{McDonald2008,Giannantonio2010,Baldauf2011}) in the expansion. The full set of terms in the Eulerian framework for an arbitrary quadratic PNG up to second order in perturbations and linear in $\fnl$ is \cite{Assassi2015}:
  
  \begin{equation}\label{eq:deltaNG}
   \delta_h^{E,(NG)}(\bx,\tau)=b_{\Psi}^E(\tau)\Psi(\bq) +b_{\Psi\delta}^E(\tau)\Psi(\bq)\delta(\bx,\tau)+\veps_{\Psi}^E(\bx,\tau)\Psi(\bq)\,,
  \end{equation}
  where $\bq$ are the spatial coordinates in the Lagrangian frame. The field $\Psi$ is a non-local transformation of the Bardeen gravitational potential, given by $\Psi(\bq)=\int d^3\bk k^{\alpha}\Phi(\bk)\exp(i\bk\bq)$, where $\alpha$ takes real values, which depend on the PNG type considered (see Ref.~\cite{Desjacques2016} for an extensive discussion). Furthermore, $\veps_{\Psi}^E$ is the stochastic counterpart of the field $\Psi$. 
  
  The bias coefficients of the general bias expansion [Eqs. (\ref{eq:deltaG}) and (\ref{eq:deltaNG})], including those of the field  $\Psi$ (\ie $b_{\Psi}$ and $b_{\Psi\delta}$) can be derived by utilising the peak-background split (PBS) argument (see \aref{app:PBSbias} for a quick review). The halo mass function, which is the mean co-moving number density of halos per logarithmic mass interval, can be parametrised in the form of \cite{PS1973}
  
  \begin{equation}\label{eq:massfun}
      n_h(M,z)=\frac{\mean{\rho}_m}{M} f(\nu)\bigg|\frac{d\ln\nu}{d\ln M}\bigg|\,,
  \end{equation}
  where $\rho_m(z)=\Omega_m(z)\rho_c^0$ is the mean co-moving density at redshift z and $\rho_c^0$ is the critical density of the Universe at $z=0$. The multiplicity function $f(\nu)$ is an arbitrary function of the peak height $\nu\equiv\delta_c/\sigma_R(M,z)$, while $\delta_c=1.686$ is the threshold value needed for a dark matter peak to form a halo and $\sigma^2_R(M,z)$ is the variance of the linear density field smoothed with a top-hat filter over a radius R and enclosed mass $M=(4\pi/3)\mean{\rho}_mR^3$. In this work, for the theoretical halo mass function we will use the best-fit results, originating from the comparison to N-body simulations, presented in Ref.~\cite{Tinker2008} (hereafter T08) and shown in \eref{eq:Tinker_MF}.
  
  For a universal mass function, like the one used here, analytic expressions can be derived for the non-Gaussian bias coefficients in Eulerian framework, after using the peak-background split bias parameters definition [\eref{eq:PBS_arg}], which are given by \cite{Desjacques2011a,Schmidt2013,Desjacques2016}:
  
   \begin{equation}\label{eq:bpsi}
     b_\Psi^E(M,z)=A\fnl\left[2\delta_cb_1^L+4\left(\frac{d\ln\sigma_{R,-\alpha}^2}{d\ln\sigma_{R}^2}-1\right)\right]\frac{\sigma_{R,-\alpha}^2}{\sigma_{R}^2},
  \end{equation}
  and \cite{Giannantonio2010,Karagiannis2018}
   \begin{equation}\label{eq:bpsidE}
   b_{\Psi\delta}^E(M,z)=2A\fnl\bigg[\delta_c\left(b_2^E+\frac{13}{21}(b_1^E-1)\right) \nonumber \\
   +b_1^E\left(2\frac{d\ln\sigma_{R,-\alpha}^2}{d\ln\sigma_{R}^2}-3\right)+1\bigg]\frac{\sigma_{R,-\alpha}^2}{\sigma_{R}^2}\,,
  \end{equation}
  where $ \sigma_{R,n}^2=1/(2\pi)^3\int d^3\bk k^n W_R(k)^2P^{L}(k,z) $, with $W_R$ being the top-hat filter, while the superscript L and E over the bias correspond to the parameters in the Lagrangian and Eulerian framework respectively. For the details of deriving Eqs. (\ref{eq:bpsi}) and (\ref{eq:bpsidE}), the reader is encouraged to check Appendix C of Ref.~\cite{Karagiannis2018} or the discussion of Ref.~\cite{Desjacques2016}. The values of $A$ and $\alpha$ depend on the PNG type considered, e.g. for the local case, $\alpha=0$ and $A=1$, these expressions reduce to the well known results $b_\Psi^E\rightarrow b_\Phi^E=2\fnll\delta_cb_1^L$ \cite{Dalal2008,Slosar2008,Giannantonio2010} and $b_{\Psi\delta}^E\rightarrow b_{\Phi\delta}^E=2\fnll[\delta_cb_2^E+(13/21\delta_c-1)(b_1^E-1)]$ \cite{Giannantonio2010,Baldauf2011,Sefusatti2012}. In the cases of equilateral and orthogonal PNG, the values are $\alpha=2$ and $A=3$ and $\alpha=1$ and $A=-3$ respectively \cite{Schmidt2010,Giannantonio2012}. Note that, for the equilateral PNG ($\alpha=2$), the scale-dependent bias correction, $b_\Psi k^\alpha/M(k)$ [\eref{eq:Z1}], approaches a constant value towards the large scales and thus becomes scale-independent. This creates strong degeneracies between $\fnle$ and the linear bias coefficient $b_1$, excluding the possibility of constraining equilateral PNG from the non-Gaussian bias term. Moreover, the transfer function present in $M(k)$, which produces a scale-dependence towards the small scales, introduces another degeneracy, this time between the PNG bias contribution and higher-order derivative bias terms. The presence of these degeneracies, on both large and small scales, indicate that the power spectrum does not essentially bear any constraining power on equilateral PNG \cite{Assassi2015}. Therefore, in the case of equilateral PNG, the forecasts coming from the two-point correlator are excluded from our analysis (see \sref{sec:RESULTS}).

   The presence of PNG introduces an additional scale-independent correction to the bias, due to the non-Gaussian correction to the mass function \cite{LoVerde2007}. In Refs.~\cite{Desjacques2009,Sefusatti2012} these offsets are derived up to the quadratic bias parameter and are given by:
   
   \begin{align}
   &\delta b_{1,NG}^E(\fnl)=-\frac{1}{\delta_c}\frac{\nu}{R_{NG}}\frac{\partial R_{NG}}{\partial\nu} \,,\\
   &\delta b_{2,NG}^E(\fnl)=\frac{\nu^2}{\delta_c^2R_{NG}}\frac{\partial^2R_{NG}}{\partial\nu^2}+2\nu(b_1^E-\frac{17}{21})\delta b_{1,NG}^E \,,
  \end{align}
  where $R_{NG}$ is the non-Gaussian correction to the mass function, as derived in Ref.~\cite{LoVerde2007}, while its amplitude is regulated by $\fnl$ (see also Appendix C of Ref.~\cite{Karagiannis2018} for the analytic results and derivation). These corrections are absorbed into $b_1$ and $b_2$ in all the expressions for brevity. Henceforth, we drop the superscript E from the bias parameters, since we consider the HI statistics at the time of observation.

\subsection{HI bias}\label{sec:HI_bias}
  
  The results of the general bias expansion (\sref{sec:BIAS}) are independent of the type of dark matter tracer \cite{Desjacques2016}. Therefore, the final ingredient in the determination of the two-point and three-point 21cm galaxy correlators is the prescription of how the neutral hydrogen populates the dark matter halos. In the spirit of the halo model \cite{Seljak2000,Peacock2000,Scoccimarro2000}, we can define the density of the neutral hydrogen, assuming that there is negligible contribution outside of the halos, as \cite{Villaescusa2014,Castorina2016}
  
  \begin{equation}
      \rho_{HI}(z)=\int n_h(M,z) M_{HI}(M,z) d\ln M.
  \end{equation}
  \noindent where $M_{HI}$ is the average neutral hydrogen mass contained inside a halo of mass M at redshift z. Hence, the abundance of neutral hydrogen will be, $\Omega_{HI}(z)=\rho_{HI}(z)/\rho_c^0$.

  Generalising the results of Ref.~\cite{Castorina2016}, where they provide the linear bias of the HI in the framework of the halo model, we retrieve the higher-order HI bias coefficients. The HI bias parameters are then given by:
  
  \begin{equation}\label{eq:bHI}
      b_{HI}^i(z)=\frac{1}{\rho_{HI}(z)}\int_0^\infty n_h(M,z) b_i^h(M,z)M_{HI}(M,z) d\ln M,
  \end{equation}
  The expressions for the local-in-matter bias parameters $b_N$ ($N\ge1$) are derived by using the T08 mass function [\eref{eq:Tinker_MF}] and the PBS argument [\eref{eq:PBS_arg}], where the details of the derivation are presented in \aref{app:PBSbias}. In the case of the linear halo bias (\ie $b_1^h$) instead of using the PBS bias results [\eref{eq:b1h}], we will use the best fit expression presented in Ref.~\cite{Tinker2010}, which achieves a better agreement with numerical results, for both low and high masses (see also Ref.~\cite{Lazeyras2016}). The analytic expressions for the higher-order halo bias (\ie $b_{i>1}^h$) derived in \aref{app:PBSbias} [Eqs.~(\ref{eq:b2h})-(\ref{eq:b4h})] were tested against simulations in Ref.~\cite{Lazeyras2016}. They show that the "PBS+T08" prediction for $b_2^h$ [\eref{eq:b2h}] deviate from their numerical results, mainly for low masses, while it still does better than the standard expression derived from the PBS argument and the mass function of Ref. \cite{ShethTormen1999} (hereafter ST99). In the case of $b_3^h$ and $b_4^h$, the PBS-derived results from the T08 mass function [Eqs. (\ref{eq:b3h}) and (\ref{eq:b4h})] are in agreement with their measurements. In order to ensure the self-consistency of this work, as well as to be congruent with the HI bias prediction of Ref.~\cite{Castorina2016}, we will use the "PBS+T08" results for the higher-order halo biases\footnote{The results for $b_3$ and $b_4$, although they are not part of the bias expansion up to second order [\eref{eq:deltaG}], are  presented due to their involvement in the theoretical error treatment presented in \sref{sec:fisher}} [\ie Eqs.~(\ref{eq:b2h})-(\ref{eq:b4h})]. 
  
  Note that, in Ref. \cite{Lazeyras2016} fitting formulas for $b_2$ and $b_3$, as a function of $b_1$, are provided. Using those, instead of the "PBS+T08" predictions, in our Fisher analysis change's the forecasts on the amplitude of PNG by few percent ($1-4\,\%$ depending the survey). The main reason for this change is the difference in the values of the bias parameters predicted by the two schemes. After the marginalization of the free parameters, which include the HI bias coefficients, the resulting $\fnl$ forecasts from the two different bias schemes will differ, due to correlations between $\fnl$ and the bias paraemters. However, the HOD recipe (described next) used here favours mass ranges, where both halo bias predictions (\ie the fitting formulas of Ref. \cite{Lazeyras2016} and those derived in Eqs. (\ref{eq:b2h}) and (\ref{eq:b3h})) are consistent with each other. This explains the small change observed on $\fnl$ forecasts, although the two halo bias schemes have different predictions. Simply put, using either of the two schemes does not have a significant impact on our Fisher forecasts.

 The final ingredient of the HI bias recipe is the relation for the neutral hydrogen mass $M_{HI}$. Here we use the fitting results of Ref.~\cite{Castorina2016}:
  
  \begin{equation}
    M_{HI}(M,z)=C(z)(1-Y_p)\frac{\Omega_b}{\Omega_m}e^{-M_{min}(z)/M}M^{\alpha(z)}
  \end{equation}
  
  \noindent where $Y_p = 0.24$ is the Helium fraction, $M_{min}$ represents the halo mass below which the HI abundance in halos is exponentially suppressed, $\alpha$ controls the efficiency of processes generating or destroying HI inside halos, and C is a normalization constant fixed by the value of $\Omega_{HI}(z)$. Note that, the HI bias parameters do not depend on the normalization of $M_{\rm HI}$. The values for the free parameters are considered to be, $\alpha=1$ and $M_{min}=5\cdot10^9 M_{sun}/h$. The resulting Eulerian bias parameters [\eref{eq:bHI}] are plotted as a function of redshift in \fref{fig:bias_params}. Alternative expressions for the neutral hydrogen mass, originating from comparisons with numerical results, are also presented in Refs. \cite{Padmanabhan2017a,Padmanabhan2017b}.

\subsection{Power spectrum and bispectrum of HI in redshift space}\label{sec:power_bisp}
  
In order to model the HI power spectrum and bispectrum in redshift space, we extend the SPT kernels (see Ref.~\cite{Bernardeau2002} for a review) to incorporate the bias expansion, discussed in \sref{sec:BIAS}, as well as the redshift space distortions (RSD) \cite{Sargent:1977,Kaiser1987,Hamilton1998}, which are treated perturbatively \cite{Verde1998,Scoccimarro1998}. The finger-of-god (FOG) effect \cite{Jackson1972} is treated phenomenologically, by introducing an exponential damping factor $D_\text{FOG}$, which characterizes the suppression of clustering power due to non-linear velocities. The analysis here is confined within the perturbative regime, where the leading order description for the redshift space power spectrum and bispectrum offers a good agreement with numerical measurements \cite{Gil-Marin:2014sta,Hashimoto:2017klo,Lazanu:2018yae,Oddo:2019run}. In the presence of a general, non-local, PNG the linear power spectrum and tree-level bispectrum in redshift space is given by (\eg see Ref~\cite{Karagiannis2018} for details):
  
     \begin{align}
     P_{HI}^s(\bk,z)&=T_b(z)^2\left[D_\text{FOG}^P(\bk,z)Z_1(\bk,z)^2P_m^L(k,z)+P_{\veps}(z)\right]+P_{\rm N}(\bk,z) \label{eq:Pgs},\\ 
   B_{HI}^s(\bk_1,\bk_2,\bk_3,z)&= T_b(z)^3\bigg\{D_\text{FOG}^B(\bk_1,\bk_2,\bk_3,z)\bigg[Z_1(\bk_1,z)Z_1(\bk_2,z)Z_1(\bk_3,z)B_{I}(k_1,k_2,k_3,z) \nonumber \\ 
   &+\Big\{2Z_1(\bk_1,z)Z_1(\bk_2,z)Z_2(\bk_1,\bk_2,z)P_m^L(k_1,z)P_m^L(k_2,z)+2~ \text{perm}\Big\}\bigg] \nonumber \\
   &+2P_{\veps\veps_{\delta}}(z)\Big[Z_1(\bk_1,z)P_m^L(k_1,z)+2~ \text{perm}\Big]+B_{\veps}(z)\bigg\}. \label{eq:Bgs} 
  \end{align}
  
  \noindent where $P_{\rm N}$ is the instrument noise (see \sref{sec:surveys}) and $T_b$ is the temperature function of the HI field. The expressions for the redshift space kernels [\ie $Z_1(\bk)$ and $Z_2(\bk_i,\bk_j)$], as well as the FOG dumping effect (\ie $D_{\rm FOG}$), can be found in \aref{app:RSD_kernels_FOG}. The redshift-space kernels depend on the line-of-sight direction $\hat{z}$, which breaks the statistical isotropy for the power spectrum and bispectrum. This means that, there is an additional angle needed to characterize the redshift space power spectrum, given by $\mu=\cos\omega=\hat{\bk}\cdot\hat{z}$. Moreover, the bispectrum is no longer characterized only by the triangle shape (i.e. the length of three wave vectors $k_1$, $k_2$ and $k_3$), but two additional variables are introduced to describe the orientation of the triangular configuration with respect to $\hat{z}$. The angle parametrization of \cite{Scoccimarro1999} is used here, where the polar angle is $\omega=\cos^{-1}(\hat{\bk}_1\cdot\hat{z})$ and the azimuthal angle is $\phi$. Then $\mu_1=\cos\omega=\hat{\bk}_1\cdot\hat{z}$, $\mu_2=\mu_1\cos\theta_{12}+\sqrt{1-\mu_1^2}\sin\theta_{12}\sin\phi$ and $\mu_3=-(k_1/k_3)\mu_1-(k_2/k_3)\mu_2$, where  $\cos\theta_{12}=\hat{\bk}_1\cdot\hat{\bk}_2$. Thus the parametrization of the redshift space bispectrum will be $B_{HI}^s(\bk_1,\bk_2,\bk_3,z)=B_{HI}^s(k_1,k_2,k_3,\mu_1,\phi,z)$. 
  
  The temperature function is given, in $\mu$K, by (see the Appendix of Ref.~\cite{StageII2018}) 
  
  \begin{equation}
       T_b=180(1+z)^2/E(z)\times4\times10^{-4}(1+z)^{0.6}
  \end{equation}
 
 \noindent where $E(z)=\sqrt{\Omega_m(1+z)^3+\Omega_k(1+z)+\Omega_\Lambda}$, for the standard dark-energy model (\ie $w_0=-1,\;w_\alpha=0$). The terms, $P_{\veps},\;P_{\veps\veps_{\delta}},\;B_{\veps}$ are generated by the presence of stochastic bias and their fiducial values are taken to be those predicted by Poisson statistics and are given by \cite{Schmidt2015,Desjacques2016}:
  
  \begin{equation}\label{eq:poisson_fid}
   P_{\veps}\equiv P_{\rm SN};\;P_{\veps\veps_{\delta}}=\frac{b_1}{2\mean{n}_{\rm eff}};\;B_{\veps}=\frac{1}{\mean{n}_{\rm eff}^2},
  \end{equation}
  where, in the HI halo model approach (\sref{sec:HI_bias}), the shot noise term is given by \cite{Castorina2016}:
  
  \begin{equation}
      P_{\rm SN}(z)=\frac{1}{\mean{n}_{\rm eff}(z)}=\frac{1}{\rho_{\rm HI}(z)}\int n_h(M,z)M_{\rm HI}^2d\ln M
  \end{equation}
    
  The tree-level redshift space bispectrum [\eref{eq:Bgs}] has additional terms, originating from the presence of PNG, of $\order{f_{\rm NL}^2}$. The fiducial value considered here for the PNG amplitude is $\fnl=0$, hence they do not contribute to the signal in a Fisher matrix forecast. Therefore, they will be neglected here in order to simplify our calculations. As mentioned in \sref{sec:BIAS}, the scale-dependent non-Gaussian correction $b_\Psi k^\alpha/M(k)$ in \eref{eq:Z1}, should be removed from the analysis in the case of equilateral PNG, for both power spectrum and bispectrum, due to strong degeneracies between $\fnle$ and other parameters on both large and small scales \cite{Assassi2015}, in order to avoid numerical contributions to the PNG signal that would be inaccessible by a LSS survey. Note that, the exclusion of large scales, due to foreground contamination (see \sref{sec:foreground}), allows us to safely neglect wide-angle and relativistic effects in the HI bispectrum. The full expression of the large-scale bispectrum, beyond any approximated treatments, is shown in Ref. \cite{Bertacca2017}.

\section{Modeling the thermal noise of 21\,cm surveys}
\label{sec:surveys}

The main difference between a galaxy survey and a line intensity mapping survey is that in the latter there is a noise component associated with the instrument itself. Such component, in most cases, dominates over the shot-noise term, which is instead the dominant noise term in traditional galaxy surveys. We model this instrumental noise component as a Gaussian noise given by \cite{Zaldarriaga2003b,Tegmark2008}
  \begin{equation}\label{eq:PSN}
    P_{\rm N}(\bk,z)=T_{sys}^2(z)\chi^2(z)\lambda(z)\frac{(1+z)}{H(z)}\left(\frac{\lambda^2(z)}{A_e}\right)^2\frac{1}{N_{pol}n_b(u)t_{\rm{survey} }}\frac{S_{area}}{\theta^2_{\rm FOV}}
 \end{equation}
for a radio intereferometer.   Here $T_{sys}=T_{sky}+T_{inst}$ is the system temperature, which is given from the sum between the sky and the instrument temperature (see Eq. D1 and D2 of \cite{StageII2018}), and $\lambda(z)$ is the redshifted wavelength of the 21cm HI line. The field-of-view is $\theta_{\rm FOV}$ and $S_{area}$ is the area of the survey in steradians, while $n_b(u)$ is the antenna distribution (\aref{app:baselines}), $N_{pol}=2$ is the number of polarizations per feed and $A_e$ is the effective beam area. Finally, $\chi$ and $t_{\rm{survey} }$, are the co-moving distance and the total observation time in hours, respectively.
 
We  consider three 21\,cm IM surveys: CHIME \cite{Newburgh2014}, HIRAX \cite{Newburgh2016} and the PUMA  survey \cite{PUMA_surv,StageII2018}. CHIME is operating as we speak, while HIRAX is in advanced proposal stage. The PUMA represents a much more aggressive concept about what will be possible in the future with a larger investment in this field. 

All these instruments are modeled somewhat simplistically. We consider realistic noise curves associated with a given distribution of baseline lengths, sky coverage and attainable system performance, but we skim over issues associated with realistic aperture synthesis due to earth rotation and angle of observation. We initially also assume a perfect phase calibration, but we will later on consider the impact of discarding the data inside the foreground wedge, to illustrate the importance of phase calibration. In short, the main goal of this paper is to investigate the scientific potential of these experiments to motivate the research into a complete control of systematics.

For the surveys PUMA and HIRAX, the field-of-view is $\theta_{\rm FOV}=\lambda(z)/D_{\rm eff}$ and the collecting area per feed is $A_e=\pi (D_{\rm eff}/2)^2$. We assume an effective dish area, $D_{\rm eff}$, due to the non-uniform illumination of the primary, given by $D_{\rm eff}^2=\eta_a D_{\rm dish}^2$, where $D_{\rm dish}$ is the physical dish size and $\eta_a=0.7$ is the aperture efficiency factor, taken to have the same value for both surveys (see Appendix D of \cite{StageII2018} for a discussion). HIRAX is assumed to be square closed packed, while PUMA is hexagonaly close packed with 50\% fill factor (i.e. a random 50\% of hexagonaly closed packed lattice sites are empty, so the array is equivalent in size to that of twice the number of elements but with quarter baseline density). For baseline density we follow the fitting formulas of \cite{StageII2018}.

In the case of CHIME, which is a cylindrical interferometer, we define the field-of-view as, $\theta_{\rm FOV}=(1.22\lambda(z)/W_{\rm cyl})\pi/2$, where $W_{\rm cyl}$ is the width of the cylinder in meters. While, the effective beam is given by, $A_e=\eta L_{\rm cyl}W_{\rm cyl}N_{\rm cyl}/N_{\rm dish}$, where the optical efficiency is taken to be $\eta=1$. In the case of the cylindrical interferometer, $N_{\rm dish}$ corresponds to the total number of feeds on $N_{\rm cyl}=4$ cylinders, while $L_{\rm cyl}$ is each cylinders length in meters. For the baseline distribution we take approximate analytical expression given in Appendix \ref{app:baselines}.

   \begin{table}
 \centering
 \begin{tabular}{lccc}
 $Parameters$ &   $CHIME$ & $HIRAX$ & $\Puma (\rm{Full}/\rm{Petite} )$\\ \hline
   redshift      &   $0.75-2.5$ & $0.75-2$  & $2-6$  \\
   packing  & packed cylinder array & square & hexagonal (50\% fill) \\
 $N_{dish}$ & $4\times 256$ & $1,024$  & $32,000/5,000$ \\
 $D_{dish}$  & $W_{cyl}=20\;m, L_{cyl}=100\;m$ & $6\;m$ & $6\;m$ \\
 $f_{\rm sky}$ & 0.6 & 0.36 & 0.5 \\
 $t_{tot}$ & $10000$\;hrs & $10000$\;hrs & $40000$\;hrs  \\
 $T_{inst}$ & $50\;K$ & $50\;K$ & $50\;K$
 \end{tabular}
 \caption{The basic specifications for three IM surveys considered here. For PUMA, the array is hexagonally close-packed}
 \label{table:srvspecs}
\end{table}

We take the total observation time to be 10,000 hours for CHIME and HIRAX \cite{Pourtsidou:2016ctq}, following their published plans. For \Puma, we have a $50\%$ filled array that will observe for 40,000 hours, which corresponds to 5 years.

The covered area for PUMA is half the sky, while for CHIME and HIRAX is $60\%$ and $36\%$ respectively. The specific values of the parameters needed to calculate the power spectrum noise are given in \tref{table:srvspecs}.

\subsection{Foreground exclusions}\label{sec:foreground}

Foregrounds are orders of magnitude brighter than the signal in 21\,cm studies \cite{Shaw:2013wza,Shaw:2014khi,Pober:2014lva,Byrne:2018dkh} and thus present an irreducible systematic. This effect makes the very small radial wavenumbers $k_\parallel=k\mu$ inaccessible \cite{Jacobson:2003wv,Furlanetto:2006jb}. However, based on our understanding of the production mechanism of the radio foregrounds, mainly composed by free-free and synchrontron emission from our galaxy and other unresolved sources, we have very good reasons to believe that they are spectrally smooth. This characteristic allows us to distinguish them from the cosmological signal, up to some value of $\kpar$, without significant losses \cite{Liu2011,Liu2012,Shaw:2013wza,Shaw:2014khi}. The exact value, bellow which the recovery is impossible, is unknown and a range of opinions has been proposed in the literature (see e.g. Refs.~\cite{Liu2011,Liu2012,Shaw:2013wza,Shaw:2014khi,Pober:2014lva}). For example, in Ref.~\cite{Shaw:2014khi} it is shown that a foreground cleaning can be achieved, making inaccessible only the modes that satisfy $\kpar<0.02\;\hoMpc$.

On the other hand, reconstruction techniques, and in particular the forward model reconstruction framework \cite{Jasche2010,Kitaura2013,Wang:2014hia,Jasche:2014vpa,Wang:2016qbz,Seljak:2017rmr,Modi:2018cfi}, has been used for IM in order to retrieve the long wavelength modes lost to the foregrounds \cite{Zhu:2016esh,Karacayli:2019iyd,Modi:2019hnu}, increasing significantly the range of scales accessible to a HI IM experiment. In particular, in Ref.~\cite{Modi:2019hnu} they manage to recover, almost perfectly, modes down to $k\simeq 0.01\;\hoMpc$. In this work, we avoid completely the foreground contaminated region by throwing away all the modes with $\kpar<\kparmin$. Following the findings of Refs.~\cite{Zhu:2016esh,Modi:2019hnu} and the suggestion of Ref.~\cite{StageII2018}, we choose an optimistic case, $\kparmin=0.01 \,\hoMpc$, and a pessimistic one, where we study the effect of increasing $\kparmin$ to $0.05 \,\hoMpc$.

In practice, there is an additional instrumental effect named foreground wedge \cite{Pober2014,Pober2015}. Wedge arises because a single interferometric baseline will see a monochromatic source away from zenith, fringing along frequency direction in exactly the same manner as non-monochromatic source at zenith. A full array with sufficiently dense coverage of the $u-v$ plane can break this degeneracy, but only if phase calibration is sufficiently accurate and stable. To date, this has not been achieved in current generation of intensity mapping and epoch of reionization arrays. Nevertheless, it is important to stress that the foreground wedge is not a fundamental astrophysical limitation, but a technical issue.

We model the wedge by removing all modes that satisfy $k_{\parallel}<k_{\rm wedge}\kperp$. The aggressiveness of the cut $k_{\rm wedge}$ is determined by the source furthest from the zenith that can corrupt the data. The most conservative assumption is that of horizon wedge, where any monochromatic source above horizon can contaminate the data. 21\,cm intensity mapping is not really competitive in this limit and therefore we do not consider this option. A more realistic modeling assumes that sources up to a certain number $N_w$ of primary beam sizes away from the zenith can have an effect, giving  \cite{Pober2014}
\begin{equation}\label{eq:wedge_prim}
    k_{\rm wedge}=\frac{rH(z)}{c(1+z)}\sin\left(1.22 N_w  \frac{\theta_{\rm FOV}}{2}\right),
  \end{equation}
  In this paper we consider values of $N_w=0,1,3$, where $N_w=0$ is the most optimistic (fully recovery of data inside the wedge) and $N_w=3$ is the most pessimistic.

\section{Fisher Matrix Analysis}\label{sec:fisher}

The Fisher matrix formalism is utilised here to predict the uncertainty on the amplitude of primordial non-Gaussianity from HI IM surveys. The approach relies on the estimation of the likelihood distribution around its maximum, which is assumed to be a multivariate Gaussian, by Taylor expanding it up to second order. The Fisher information matrix is then given by the second derivative of the log-likelihood $\mathcal{L}$ with respect to the parameters $\mathbf{p}$ of the assumed model by:

\begin{equation}
    F_{\alpha\beta}=-\left\langle\,\frac{\partial^2\ln \mathcal{L}}{\partial p_\alpha\partial p_\beta}\,\right\rangle,
\end{equation}
where $\mathbf{p}$ is the parameter vector and $\alpha$,$\beta$ are the indices of the vector that correspond to the unknown parameters of interest. The inverse of the Fisher matrix $(F^{-1})_{\alpha\beta}$ yields an estimate of the parameter covariance, with the smallest possible uncertainty on the measurement errors (\ie the Cramer-Rao bound). If all the free parameters of the model are measured simultaneously, we can derive the marginalised error on the parameter $p_\alpha$, i.e. taking into account degenerecies and cross-correlations between the parameters on the final constrains, by  $\sigma(p_\alpha)=\sqrt{(F^{-1})_{\alpha\alpha}}$.

 In the case of the HI power spectrum, the Fisher matrix is given by
    \begin{equation}\label{eq:fisherPs}
    F_{\alpha\beta}^{Ps}(z)=\sum_k\int_{-1}^1 \frac{d\mu}{2} \frac{\partial P_{HI}^s(\bk,z)}{\partial p_{\alpha}}\frac{\partial P_{HI}^s(\bk,z)}{\partial p_{\beta}}\frac{1}{\Delta P^2(\bk,z)} \, ,
   \end{equation}
   
   \noindent while for the bispectrum we have \cite{Scoccimarro2003}
   
   \begin{equation}\label{eq:fisherBs}
    F_{\alpha\beta}^{Bs}(z)=\frac{1}{4\pi}\sum_T\int_{-1}^{1}d\mu_1\int_0^{2\pi}d\phi\,\frac{\partial B_{HI}^s(\bk_1,\bk_2,\bk_3,z)}{\partial p_\alpha}\frac{\partial B_{HI}^s(\bk_1,\bk_2,\bk_3,z)}{\partial p_\beta}\frac{1}{\Delta B^2(\bk_1,\bk_2,\bk_3,z)}\, ,
   \end{equation}
   where the derivatives, for both correlators, are evaluated at the fiducial value of the parameter vector $\mathbf{p}$ [\eref{eq:params}]. The sum over the triangles is denoted as $\Sigma_T$, where after applying the triangle condition, we take into consideration the symmetry under all permutations of $\bk_1$, $\bk_2$ and $\bk_3$ by restricting the summation of the magnitudes to $k_{\rm min}\le k_3\le k_2\le k_1\le k_{\rm max}$.  The bin size of the wavenumber magnitude is $\Delta k$, which is taken to be the fundamental frequency of the corresponding survey, $k_f= 2\pi/L$, where we approximate for simplicity the survey volume as a cube, $L=V^{1/3}$. This bin size is used, not only for the triangle summation in \eref{eq:fisherBs}, but also for the wavenumber sum in \eref{eq:fisherPs}. The wavenumber is binned between a minimum value $k_{\rm min}=k_f$, which is the largest scale available to each survey, and a maximum value $k_{\rm max}$, which corresponds to the smallest scale assumed.
   
   In our Fisher matrix analysis, only the diagonal part of the covariance matrix (\ie $\Delta P^2$ and $\Delta B^2$) is taken into consideration, neglecting all the cross-correlations between different triangles (bispectrum) and $k$-bins (power spectrum). We adopt a Gaussian approximation for the variance terms \cite{Sefusatti2006,Sefusatti2007}:

  \begin{align}
    &\Delta P^2(\bk,z)=\frac{4\pi^2}{V_{\text{survey}}(z)k^2\Delta k(z)}P_{HI}^s(\bk,z)^2\, , \label{eq:deltaP2} \\
    &\Delta B^2(\bk_1,\bk_2,\bk_3,z)=s_{123}\pi V_f(z)\frac{P_{HI}^s(\bk_1,z)P_{HI}^s(\bk_2,z)P_{HI}^s(\bk_3,z)}{k_1k_2k_3\Delta k(z)^3}\, ,\label{eq:deltaB2}
   \end{align}

 \noindent where $s_{123}=6,2,1$ for equilateral, isosceles and scalene triangles respectively. The volume of the fundamental shell in Fourier space is  $V_f=k_f^3$. Higher order correlators are involved in the calculations of the off-diagonal terms of the covariance, for both the power spectrum (4-point) and the bispectrum (6-point) \cite{Sefusatti2006}, which makes their implementation into the Fisher analysis very challenging. In Ref.~\cite{Chan2017} they study the halo power spectrum and bispectrum covariances by comparing to N-body simulations. In the case of the power spectrum, they demonstrate that for dense samples and the redshift ranges, like those considered here, the off-diagonal terms of the covariance can be safely neglected. Furthermore, for the bispectrum, they show that the off-diagonal part can be important for low density samples and at low redshifts, while for scales $k\gtrsim0.1\;\hoMpc$ these elements significantly deviate the covariance from the Gaussian approximation. However, for the samples, redshift ranges and scales considered in this work, we do not expect the exclusion of the off-diagonal part of the covariance to have a significant impact to our final PNG forecasts. On the other hand, higher-order corrections to the bispectrum variance (i.e. the diagonal part of the covariance), could have a significant effect not only for cosmological parameters \cite{Chan2017}, but also for PNG constraints \cite{Karagiannis2018}. The full non-Gaussian contribution to the variance, for the large scales considered in this work, can be well approximated by perturbative corrections to the power spectrum appearing in \eref{eq:deltaB2}, obtaining \cite{Chan2017}:
 
\begin{equation}\label{eq:DB2_NL}
 \Delta B_\text{NL}^2(\bk_1,\bk_2,\bk_3,z)=\Delta B^2(\bk_1,\bk_2,\bk_3,z)+s_{123}\pi V_f(z)\frac{P_{HI}^s(\bk_1,z)P_{HI}^s(\bk_2,z)P_{HI}^\text{NL}(\bk_3,z)+2\perm}{k_1k_2k_3\Delta k(z)^3} \, ,
\end{equation}
 where $P_{HI}^\text{NL}(\bk,z)$ is given by \eref{eq:Pgs} after substituting the linear power spectrum with the nonlinear correction, i.e. $P_m^L(k)\rightarrow P_{HI}^\text{NL}(k)-P_m^L(k)$, where the nonlinear power spectrum $P_{NL}^\text{NL}$ is taken to be the HALOFIT power spectrum \cite{Smith2003,Takahashi2012}. For the bispectrum Fisher matrix forecasts [\eref{eq:fisherBs}], the variance with the non-Gaussian correction [\eref{eq:DB2_NL}] will be used.
 
  The set of free parameters that consist the parameter vector, used in the derivatives of the Fisher matrix formalism, is considered to be
  
      \begin{equation}\label{eq:params}
    \mathbf{p}=\{\fnl,b_1,b_2,b_{s^2},P_{\veps},P_{\veps\veps_{\delta}},B_{\veps},f,\sigma_{\upsilon}\}.
   \end{equation} 
   where $\sigma_{\upsilon}$ controls the strength of the FOG dumping factor (see \aref{app:RSD_kernels_FOG} for the details).The stochastic bias contributions (\ie $P_{\veps},P_{\veps\veps_{\delta}}$ and $B_{\veps}$) are considered here as nuisance parameters and they are marginalised over to acquire the Fisher sub-matrix for the parameters of interest, i.e. $\fnl,b_1,b_2,b_{s^2},f,\sigma_{\upsilon}$. After acquiring the sub-matrix we marginalise over the rest of the free parameters for each redshift slice (see e.g. \cite{Wang2006} for details on the process). Here the redshift slices are considered non-overlapping (i.e. cross-redshift correlation are considered to be zero). We then proceed to the summation of the final $F_{\fnl\fnl}(z)$ over the whole redshift range of each survey, in order to derive the final forecasts on the amplitude of PNG from each respective correlator, i.e. $F_{\alpha\beta}^{Ps}$ and $F_{\alpha\beta}^{Bs}$. We also consider the forecasts from the combined power spectrum and bispectrum Fisher matrices, i.e. $F_{\alpha\beta}^{Ps+Bs}=F_{\alpha\beta}^{Ps}+F_{\alpha\beta}^{Bs}$, where we neglect the cross-covariance between the two. This would have minimal impact on our forecasts \cite{Chan2017}. Note that, cosmological parameters are considered to be know (fixed), since they can be measured with high accuracy by other probes (e.g. CMB, BAO etc.). For the CMB primordial bispectrum, it was shown in Ref.~\cite{Liguori2008} that the degeneracies between cosmological parameters and $\fnl$ are small, therefore we do not expect that the propagation for the cosmological errors will significantly impact the $\fnl$ constraints. This was also pointed out, for the case of the galaxy power spectrum and bispectrum, in Refs.~\cite{Giannantonio2012,Moradinezhad2018,Bellomo:2020pnw}.
   
 The Fisher analysis is confined inside the perturbative regime, where the SPT description of the power spectrum and bispectrum is reliable (for the latter see e.g. Refs.~\cite{Gil-Marin:2014sta,Hashimoto:2017klo,Lazanu:2018yae,Oddo:2019run,Agarwal:2020lov}), by cutting the smallest scales at $k_{max}=0.75 k_{NL}(z)$. The non-linear scales $k_{NL}$ are taken to be the linear, one dimensional velocity dispersion, given by

\begin{equation}\label{eq:knl}
k_{\rm NL}(z) =\left[ \frac{1}{6\pi^2} \int_0^\infty dk\, P_{\rm lin}(k,z)  \right]^{-1/2} \ .
\end{equation}
The increase of the maximum scales with redshift takes into advantage the fact that the Universe is more linear at higher redshifts, hence the tree-level description of the bispectrum holds up to higher values of $k_{\rm max}$. This means that we can include more modes into the analysis for an increasing redshift, which will enhance the number of formed triangles and therefore the bispectrum signal (i.e. the information of the Fisher matrix) through the reduction of the cosmic variance. 

Nonetheless, ignoring higher-to-leading order terms (i.e. 1-loop) in the modelling can affect the forecasts. In a perturbative approach each order has a limited range of validity and the higher order contributions could become important, while approaching the nonlinear regime. Although the analysis is limited up to linear scales and therefore the higher order corrections are not expected to significantly affect our forecasts, we take into account the parameter shift due to neglecting the 1-loop contributions at the level of the Fisher matrix, through the theoretical errors approach \cite{Baldauf2016}. In this formalism the theoretical errors are defined as the difference between the chosen perturbative order (i.e. tree-level) and the next higher order (i.e. 1-loop). An envelope $\mathbf{E}$ is fitted, bounding these errors, as a function of the wavenumber. Together with a correlation coefficient $\rho_{ij}$, which is assumed to be a multivariate Gaussian, we form a theoretical error covariance $C_{ij}^e=E_i\rho_{ij}E_j$, where $i,\;j$ are the indices of the different momentum configurations (i.e. number of bins and triangles for the power
spectrum and bispectrum respectively). The correlation coefficient takes into account the correlations between the different momentum configurations, making the impact of the theoretical errors independent of the k binning and the correlation length (see Ref.~\cite{Baldauf2016} for an extensive discussion). The final covariance used in the Fisher matrix analysis [Eqs. \eqref{eq:fisherPs} and \eqref{eq:fisherBs}] will be the sum of the error covariance $C^e$ with the respective variance [i.e. Eqs. \eqref{eq:deltaP2} and \eqref{eq:DB2_NL}]. The envelopes, for the power spectrum and bispectrum modelling, are taken from Ref.~\cite{Karagiannis2018}, where they have extended the approach of Ref.~\cite{Baldauf2016} to include the theoretical uncertainties from excluding the 1-loop terms of the matter and the local-in-matter bias expansions (i.e. $b_1,\;b_2,\;b_3$, etc.) in the power spectrum and bispectrum, while performing an extensive study on the impact of theoretical errors on various parameters, including $\fnl$.

Note that the smallest accessible scales are also limited by the specifications of each survey.  The largest and smallest available perpendicular scales are given by $k_{\bot ,min}=2\pi /(\chi \theta_{\rm FOV})$ and $k_{\bot ,max}=2\pi D_{\rm max}/(\lambda \chi)$ respectively. These limits originate from the intrinsic limitations of interferometers, which are fundamentally unable to probe any scale larger than those that correspond to their dish diameter \cite{Bull2015}. All these scale limits, i.e. the foreground cut off ($\kparmin$), the wedge [\eref{eq:wedge_prim}], the intrinsic cut offs of the interferometer and the scale of model validity ($k_{\rm max}$), constitute an observational window, where the signal of the HI power spectrum and bispectrum, in the redshift range considered, in non-zero. The signal from this window accounts for the forecast results presented in the next section. Note, that the window becomes more narrow as we go to higher redshifts, due to the increasing contamination, which for the case of the HI bispectrum leads to a significant reduction in the signal (see \fref{fig:base}). For redshifts of the Epoch of Reionization, the window for the bispectrum could close completely and an aggressive foreground cleaning formalism would be needed \cite{Watkinson:2020zqg}.

\section{Results}\label{sec:RESULTS}

Our main results are summarized in Table \ref{table:forc_all}. We give results for all three types of non-Gaussianity and for a number of wedge cuts. For comparison, the currently achieved sensitivities from Planck \cite{Planck_PNG2016} are 5, 43 and 21 for squeezed, equilateral and orthogonal shapes respectively. Forecasted numbers for CMB-S4 experiments are 2, 21 and 9. SphereX can achieve 0.5 for the local shape. Galaxy surveys are generally unable to achieve competitive constraints for shapes which peak away from the squeezed limit, due to heavy contamination from non-linear gravitational effects; this is particularly true for the equilateral case, where late-time non-linear contributions are largest.

\begin{table*}
\centering
\resizebox{\linewidth}{!}{
\begin{tabular}{c|ccc|ccc|ccc|}
\cline{2-10}
                                & \multicolumn{3}{|c|}{CHIME} & \multicolumn{3}{c|}{HIRAX} & \multicolumn{3}{c|}{PUMA \rm{Full (Petite)}} \\ \hline
                           \multicolumn{1}{|c|}{Wedge type} & {NO} & {PB} & {$3\times$ PB} & {NO} & {PB} & {$3\times$ PB} & {NO} & {PB} & {$3\times$ PB} \\ \hline                           
\multicolumn{1}{|c|}{P(loc)}    & 31.7 & 31.7 & 31.9 & 25.3 & 25.3 & 25.8 & 1.61 (1.67) & 1.72 (1.77) & 2.52 (2.57)           \\
\multicolumn{1}{|c|}{B(loc)}     & 71.9 & 71.9 & 72.7 & 9.2 & 9.2 & 10.2 & 0.31 (0.77) & 0.41 (0.91) & 0.91 (1.67)                \\
\multicolumn{1}{|c|}{P+B(loc)}   & 28.2 & 28.2 & 28.4 & 8.4 & 8.5 & 9.3 & 0.3 (0.69) & 0.4 (0.8) & 0.84 (1.37)               \\ \hline
\multicolumn{1}{|c|}{P(equil)}   & - & - & - & - & - & - & - &- &-                       \\
\multicolumn{1}{|c|}{B(equil)}   & 569.9 & 569.9 & 576.7 & 98.0 & 98.3 & 112.5 & 11.66 (21.87) & 16.71 (27.09) & 41.99 (55.49)               \\
\multicolumn{1}{|c|}{P+B(equil)} & 257.3 & 257.3 & 259.8 & 51.7 & 51.9 & 59.1 & 5.06 (10.15) & 7.98 (13.29) & 23.17 (29.38)              \\ \hline
\multicolumn{1}{|c|}{P(ortho)}   & 937.9 & 937.9 & 940.3 & 613.6 & 613.6 & 662.4 & 39.41 (45.29) & 46.94 (51.71) & 74.97 (78.96)               \\
\multicolumn{1}{|c|}{B(ortho)}   & 215.3 & 215.3 & 216.7 & 34.9 & 35.0 & 38.4 & 3.13 (7.33) & 4.22 (8.52) & 8.86 (14.28)              \\
\multicolumn{1}{|c|}{P+B(ortho)} & 158.2 & 158.2 & 159.5 & 28.5 & 28.5 & 31.8 & 3.04 (6.53) & 4.08 (7.79) & 8.56 (13.32)                \\ \hline
\end{tabular}
}
\caption{Forecasts for the $1-\sigma$ error of the primordial non-Gaussian amplitude in the case of the three PNG types considered here. These results come from the summation of the signal over the whole redshift range of each survey, where we show the constraints originating from the galaxy power spectrum, bispectrum and their combined signal. In addition, different wedge cases are considered for each survey. The forecasts under the column "NO" correspond to the case where no wedge cut is used, while under the columns titled "PB" and "3xPB" are the forecasts after applying the prime-beam wedge cuts [\eref{eq:wedge_prim}] for $N_w=1$ and $N_w=3$ respectively. Moreover, we exclude from the analysis all scales that satisfy, $k_\parallel< 0.01\;{\rm h/Mpc}$, as discussed in the main text. The forecasts under "3xPB" correspond to the main results of this work.}
\label{table:forc_all}
\end{table*}

We see that CHIME is never competitive, despite significant volume coverage, because the thermal noise is overwhelming. HIRAX could instead achieve similar sensitivities to Planck. These constraints will be independent and could therefore improve over Planck by some 40\%. Most importantly, if such result will be achievable in practice -- keeping all systematics under control -- it will provide an outstanding observational confirmation that primordial non-Gaussianity is indeed a very promising field of study for future, ambitious intensity mapping surveys.
Our forecasts show that PUMA should be competitive with CMB-S4 and Spherex. Notably, it could provide particularly strong constraints for the equilateral shape, if the foreground wedge could be controlled. This is of particular interest, considering the general difficulty of improving equilateral constraints using LSS tracers, as mentioned above.

\begin{figure*}
\centering
\resizebox{\linewidth}{!}{\includegraphics{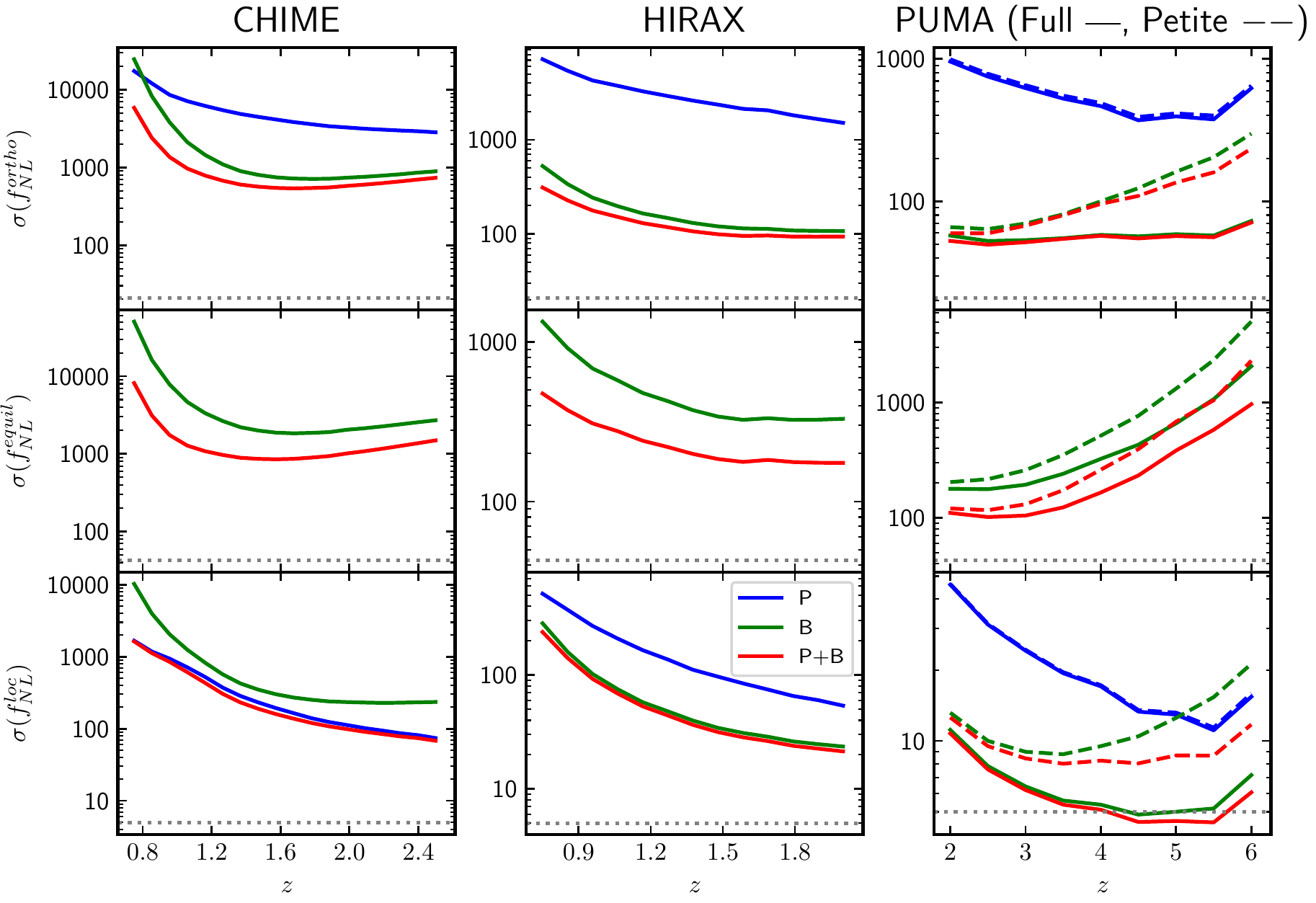}}
    \caption{The forecasts for the non-Gaussian amplitude, in the case of the three types of PNG considered here, after marginalising over the free parameters. The results from the powerspectrum (blue), bispectrum (green) and their combined signal (red) are plotted as a function of redshift, while each column corresponds to one of the three surveys considered in this work. Each point of the lines corresponds to a fixed redshift slice with $dz=0.1$. The ''$3\times$prime-beam" wedge is used, as well as we exclude all scales that satisfy, $k_\parallel<0.01\;{\rm h/Mpc}$. The dotted grey line indicates the best constraints on the PNG amplitude, as given by Ref.~\cite{Planck_PNG2016}.} \label{fig:base}

\end{figure*}

In \fref{fig:base} we study where the information is coming from, as a function of redshift. In general, the neutral hydrongen maps are noisier at higher redshift, due to increasing sky noise temperature and decreasing bandwidth per comoving distance. This is partly offset by the fact that the non-linear scale is smaller at higher redshift (\ie increasing $k_{NL}(z)$), and that the total volume per sky area is larger.  As a result, there are no unique trends and different experiments can extract most information from either low or high redshift end, depending both on the experiment and on the type of non-Gaussianity under consideration.

In most cases, presented in \fref{fig:base}, the majority of the PNG signal originates from the bispectrum (see also \tref{table:forc_all}), highlighting the importance of three-point statistics in constraining PNG from future IM experiments, and in particular from packed interferometric arrays. At higher redshift slices the Universe becomes more linear, increasing the scale range where the linear theory is still valid (increasing $k_{\rm max}$). A boost in the constraining power of the bispectrum is hence expected, due to the growing number of formed triangles and the amplitude reduction of the gravitational contaminants. This can be observed in some cases  shown in \fref{fig:base}, but a trend cannot be established due to the presence of observational and other effects (see \sref{sec:foreground}), as well as due to the individual traits of the PNG types considered.

In the case of local PNG, the bispectrum signal peaks on the squeezed configurations ($k_1 \ll k_2 \sim k_3$), therefore sufficiently large and small (up to the validity of linear theory) scales must be accessible by a survey, in order to have enough squeezed triangles to produce compelling bispectrum constraints. If they are restricted, due to \eg observational and instrumental effects or low redshift slices (\ie smaller linear regime), then the scale-dependence in the galaxy power spectrum provides most of the PNG signal. This is the case for CHIME and PUMA Petite, where for the latter this is evident for the large redshift slices (\ie $z\ge 5$).  

In the equilateral PNG scenario, the scale-dependent bias term approaches a constant value on large scales. In addition, as was discussed before, the presence of degeneracies between $\fnle$ and other parameters \cite{Assassi2015}, strip power spectrum from essentially any constraining power on equilateral PNG. This leaves galaxy bispectrum to be the sole contributor of the signal. The increasing range of the linear regime (\ie increase in the number of the formed equilateral triangles) with redshift, in the cases of CHIME and HIRAX, improves the constrains on $\fnle$, up to a saturation point due to the applied scale cuts (see \sref{sec:foreground} and \sref{sec:fisher}). These scale limitations are the ones responsible for the opposite trend observed in the PUMA results. Due to the pessimistic wedge cuts, as well as the $k_{\rm min,||}$ and $k_{\max}$ limits, the expected wide scale range, accessible to PUMA, is shrunken towards high redshift slices, rendering their contribution to the integrated signal minimal. The same behaviour is observed for orthogonal PNG, where the effect of the k-cuts can be now seen also in the power spectrum constraints. Due to the functional form of the orthogonal PNG scale-dependent bias \cite{Desjacques2011a,Schmidt2013}, only the very high redshift slices of PUMA are affected by the scale cuts.

\begin{figure*}
\centering
\resizebox{\linewidth}{!}{\includegraphics{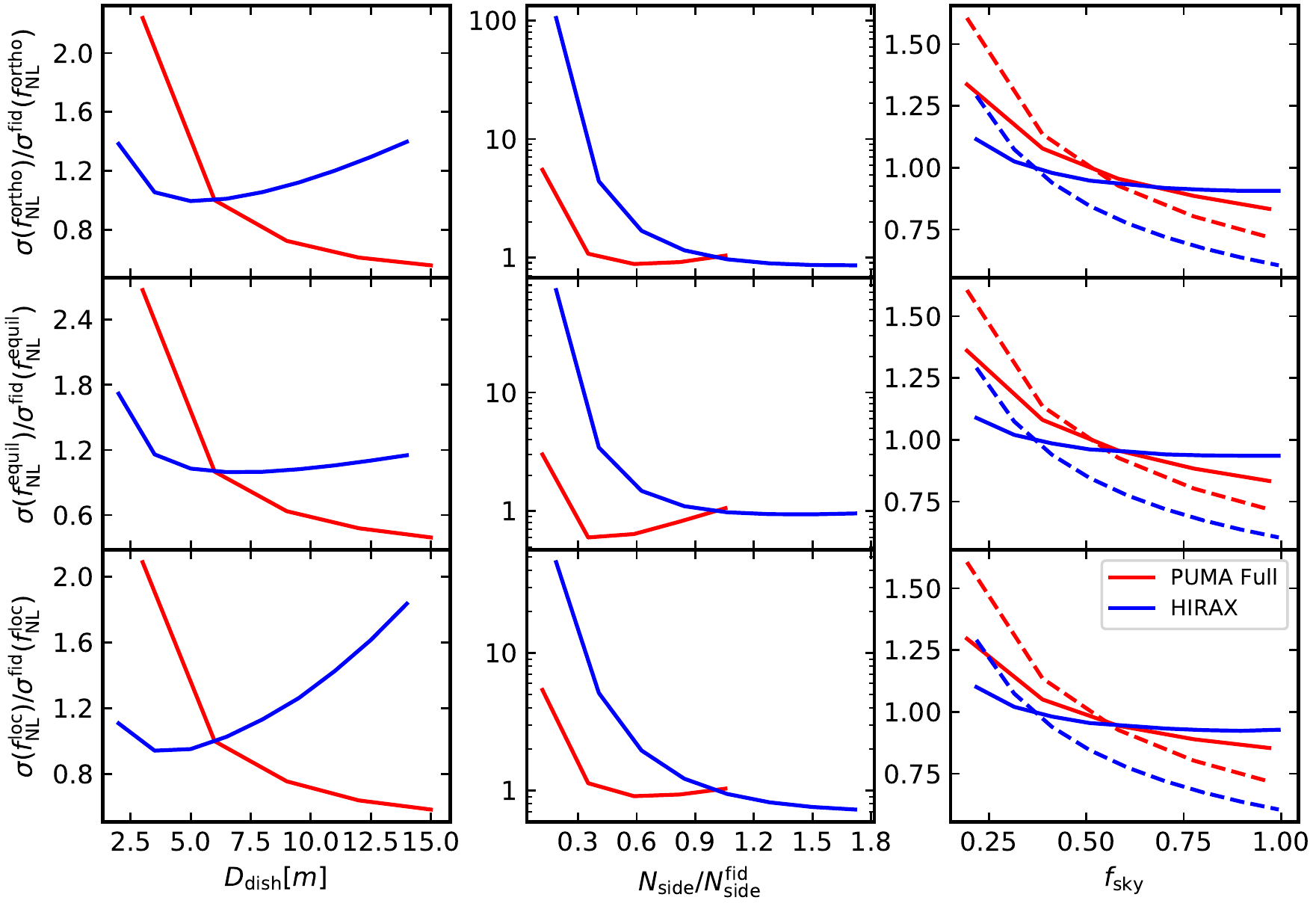}}
    \caption{The PUMA (red) and HIRAX (blue) forecasts on $\fnl$, normalised over the fiducial forecasts (see \tref{table:forc_all}), for the three PNG shapes considered here, as a function of the dish size (lefts panels), the number of dishes in each side (middle panels) and $f_{\rm sky}$ (right panels). All the remaining parameters (including integration time) are kept fixed for all three cases, beside the $N_{\rm side}$ case (middle panels), where in addition we keep fixed the collecting area. These results correspond to the combined power spectrum and bispectrum signal, after summing over the whole redshift range of the survey. The prime-beam wedge for $N_w=3$ is used, as well as all scales that satisfy, $k_\parallel<0.01\;{\rm h/Mpc}$, are excluded. The plotted dashed lines in the right panels represent the function, $(f_{\rm sky}/f_{\rm sky}^{\rm fid})^{-1/2}$. Note that, in the middle panels the results for PUMA reach up to $N_{\rm side}/N_{\rm side}^{\rm fid}\sim 1.1$ due to the tremendous increase in the calculation time. Nonetheless, the significant part of the functional dependence is shown.} \label{fig:sfnl_specs_PUMA}

\end{figure*}

In \fref{fig:sfnl_specs_PUMA} we show how experimental design parameters affect constraints for HIRAX and PUMA. We consider several changes. Changing the dish size (keeping other parameters fixed), moves sensitivity wholesale towards smaller perpendicular scales, affecting the minimum available $\kperp$, but also raising the higher $\kperp$ available. The increase of the dish size decreases the thermal noise, due to the functional form of the baseline distribution \footnote{Note that, this would not be the case, if a uniform baseline distribution had been used.} [\eref{eq:base_SII_XIR}], while the value of the wedge cutoff $k_{\rm wedge}$ is also decreased due to the dependence on $\theta_{\rm FOV}$, allowing more modes to be used in the Fisher analysis. The latter is particularly important for the PNG signal originating from the high redshifts, especially those accessible to PUMA, since the contaminated region is increased with redshift due to the dependence on the comoving distance and the Hubble parameter [\eref{eq:wedge_prim}]. Therefore we find that PUMA prefers bigger dishes, while for HIRAX the 6m design size is nearly optimal (left panels in \fref{fig:sfnl_specs_PUMA}). Second, changing the number of dishes at fixed collecting area (middle panels in \fref{fig:sfnl_specs_PUMA}) lowers the noise power spectrum while keeping the resolution the same. We parametrize this with a parameter $N_{\rm side}=\sqrt{N_{\rm dishes}}$, which reduces to just the side of the array measured in the number of dishes for HIRAX. 
Naively, one would expect that increasing $N_{\rm side}$ can only improve the results, until the sample variance starts to dominate and the results converge to volume limited measurement. However, we find that this is not the case for PUMA, where increasing $N_{\rm side}$ actually makes things worse, especially for the equilateral case. This is because this exercise is done at the $N_{\rm w}=3$ wedge cut. Increasing the number of dishes at fixed collecting area makes dishes smaller, exacerbating the wedge cut. As a result, PUMA prefers larger dishes at fixed collecting area, which is consistent with our previous observation, i.e. after changing the dish size (left panels of \fref{fig:sfnl_specs_PUMA}).  In the final column we plot trends with the observed sky area. This is done at a fixed observation time, so the smaller $f_{\rm sky}$ implies a smaller fraction of the sky, but observed with a lower noise level. We would therefore expect the error bars to follow
\begin{equation}
  \sigma \propto \left(\frac{P_S + P_{\rm SN}+ \frac{f_{\rm sky}}{f_{\rm sky,fid}} P_N}{f_{\rm sky}} \right)^{-1/2}.
\end{equation}
In other words, for the sample variance limited case, we expect the sensitivity to improve as $f_{\rm sky}^{-1/2}$ (plotted as dashed lines) and for the thermal noise dominated case we expect curves to flatten. We find that HIRAX indeed flattens out, but that PUMA is somewhere between the two limits.

Finally in Table \ref{table:forc_kparmin} we study the effect of changing the irreducible value of foreground filtering $\kparmin$. Since foregrounds affect mostly large angular scales, we find that the effect is the largest for the squeezed triangle configurations, which contain one small side. This leads to a significant decrease in the constraining power of both power spectrum and bispectrum in the local PNG case. The reduction is still very pronounced for orthogonal configurations. This is not completely intuitive, but explainable with the fact that the orthogonal shape presents a non-negligible correlation with the squeezed one, and takes significant contributions from flattened triangles. Equilateral PNG is the least affected, as expected, since only a small amount of equilateral triangles, formed by the largest scales, are excluded.

\begin{table*}
\centering
\resizebox{0.8\linewidth}{!}{
\begin{tabular}{c|cc|cc|cc|}
\cline{2-7}
                                & \multicolumn{2}{|c|}{CHIME} & \multicolumn{2}{c|}{HIRAX} & \multicolumn{2}{c|}{PUMA Full} \\ \hline
                           \multicolumn{1}{|c|}{$k_{\parallel, min}\;[{\rm h/Mpc}]$} & 0.01 & 0.05 & 0.01 & 0.05 & 0.01 & 0.05 \\ \hline                           
\multicolumn{1}{|c|}{P(loc)}     & 31.9 & 105.5 & 25.8 & 101.3 & 2.52 & 8.42             \\
\multicolumn{1}{|c|}{B(loc)}     & 72.7 & 457.7 & 10.2 & 71.5 & 0.91 & 3.63              \\
\multicolumn{1}{|c|}{P+B(loc)}   & 28.4 & 101.7 & 9.3 & 47.9  & 0.84 & 3.05             \\ \hline
\multicolumn{1}{|c|}{P(equil)}   & - & - & - & - & - & -   \\
\multicolumn{1}{|c|}{B(equil)}   & 576.7 & 3139.0 & 112.5 & 484.5 & 42 & 77.88              \\
\multicolumn{1}{|c|}{P+B(equil)} & 259.8 & 777.4 & 59.1 & 122.1 & 23.17 & 30.38             \\ \hline
\multicolumn{1}{|c|}{P(ortho)}   & 940.3 & 1502.9 & 662.4 & 906.6 & 74.97 & 80.32              \\
\multicolumn{1}{|c|}{B(ortho)}   & 216.7 & 1242.5 & 38.4 & 203.8 & 8.86 & 30.24            \\
\multicolumn{1}{|c|}{P+B(ortho)} & 159.5 & 600.5 & 31.8 & 102.3 & 8.56 & 24.25              \\ \hline
\end{tabular}
}
\caption{ Same as in \tref{table:forc_all}, but now we test the effect of the $\kparmin$ cuts on the $\fnl$ forecasts. We fix the wedge cuts to be as in the case of ''$3\times$ prime-beam" (\ie $N_w=3$ in \eref{eq:wedge_prim}) and consider two values of the $k_\parallel$ cut. Therefore the results that are under the "$k_{\parallel,min}=0.01$" columns correspond to the main results of this work (\ie same as "$3\times$PB" of \tref{table:forc_all}).}
\label{table:forc_kparmin}
\end{table*}

\section{Conclusions}\label{sec:CONCL}

In this paper we have studied constraints on primordial non-Gaussianity from present and future 21\,cm intensity mapping experiments. Since these experiments cover huge volumes of space, it is intuitive that they should be able to produce competitive non-Gaussianity results.

We find that a futuristic experiment such as PUMA will produce very competitive non-Gaussianity bounds from purely internal measurements of the power spectrum and bispectrum (see \tref{table:forc_all}). This applies in particular to orthogonal and equilateral shapes, generally very hard to constrain using galaxy surveys, due to heavy contamination from non-linear late time evolution of structures. In the case of PUMA, this issue is offset by large volume coverage and nearly sample variance limited measurements. Moreover, at higher redshift, where the non-linear scale is smaller (larger $k_{\rm NL}$), the increasing number of formed triangles leads to a growth in the bispectrum signal. Additionally the amplitude reduction of the gravitational contaminants, leads to a promising PNG signal for high redshift surveys. In the case of PUMA, we find that the presence of foreground wedge significantly worsens the limits derived from the high redshift slices (see \fref{fig:base}). It is therefore imperative that this important systematic is brought under control.

We find that constraints on the local shape of bispectrum are tight, but considerably less impressive compared to other probes of non-Gaussianity, such as those coming from LSST \cite{Karagiannis2018} or the upcoming SPHEREx experiment \cite{Dore2014}. This is because the foreground cut at $\kparmin=0.01\hoMpc$ leads to loss of information on large scale modes, which are required in the squeezed limit. While PUMA on its own is not competitive, cross-correlations between the small scale two-point function from PUMA and large scale mode measured by other means (traditional galaxy survey or CMB lensing) could yet prove to be very effective in measuring local non-Gaussianity (see Refs.~\cite{Alonso:2015sfa,Fonseca2015} for a multi-wavelength power spectrum application). We leave this work for the future.

We find that the current and upcoming generation of 21\,cm surveys, such as CHIME and HIRAX are in general not very competitive. This boils down mainly to a considerable thermal noise in the field measurements. Any departure from sample variance limit hits the bispectrum measurement more than the power spectrum measurement and this leads to relatively non-competitive numbers in this case. 

We also find that the difference between the full and petite versions of PUMA is pretty modest in the case of pessimistic foreground, but can be considerable if no wedge is assumed.  The main difference between these two experiments is in angular resolution, as the full PUMA is simply an extended version of PUMA Petite. This leads to both, additional information on smaller scales but also improved sampling of large scales, since the number of short baselines increases too. What we see is a complex interplay between the change in noise and increase in the number of possible triangle configurations, that depend on both the maximum wavenumbers, but also any other cuts on the Fourier plane. The bottom line is that minimizing the effect of the foreground wedge is essential in order to extract all possible science from these measurements. 

In traditional radio astronomy intuition, splitting the same collecting area into more interferometric elements always improves results, because it increases the field of view at constant noise. We find, however, that wedge considerations favor larger dishes, even at the same total collecting area. Since dishes can be thought of as analog interferometers, this in effect boils down to the relative difficulty of analog phase calibration (i.e. surface accuracies) vs electronic phase calibration. The actual trade off calculation will require more sophisticated studies of relative contributions of individual element repeatability and stability, inteferometric  phase control and data reduction software imperfections to the total error. 

To conclude, measurements of primordial non-Gaussianity are ideally suited to intensity mapping of 21\,cm neutral hydrogen across cosmic epochs, provided the thermal noise can be made smaller with a sufficiently large array and assuming systematics can be brought under control. Stage {\sc II} 21\,cm experiments such as PUMA will be therefore able to make impressive measurements of non-Gaussianity.

\[\]{\bf Acknowledgments:}\\
The authors would like to thank Emanuele Castorina for useful comments on the draft of the manuscript, as well as Titouan Lazeyras for helpful feedback on the fist arXiv version. ML's work is supported by the University of Padova under the STARS Grants programme CoGITO, Cosmology beyond Gaussianity, Inference, Theory and Observations. ML acknowledges support from the INDARK INFN Initiative (web.infn.it/CSN4/IS/Linea5/InDark).

\bibliography{bibliography}

\providecommand{\href}[2]{#2}\begingroup\raggedright\begin{thebibliography}{100}

\bibitem{2019BAAS...51c.107M}
P.~D. {Meeburg}, D.~{Green}, R.~{Flauger}, B.~{Wallisch}, M.~C.~D. {Marsh},
  E.~{Pajer} et~al., \emph{{Primordial Non-Gaussianity}}, {\emph{\baas}
  {\bfseries 51} (2019) 107}
  [\href{https://arxiv.org/abs/1903.04409}{{\ttfamily 1903.04409}}].

\bibitem{1412.4671}
M.~{Alvarez}, T.~{Baldauf}, J.~R. {Bond}, N.~{Dalal}, R.~{de Putter},
  O.~{Dor{\'e}} et~al., \emph{{Testing Inflation with Large Scale Structure:
  Connecting Hopes with Reality}}, {\emph{arXiv e-prints} (2014)
  arXiv:1412.4671} [\href{https://arxiv.org/abs/1412.4671}{{\ttfamily
  1412.4671}}].

\bibitem{Bartolo2004b}
N.~{Bartolo}, S.~{Matarrese} and A.~{Riotto}, \emph{{Non-Gaussianity in the
  curvaton scenario}},
  \href{https://doi.org/10.1103/PhysRevD.69.043503}{\emph{\prd} {\bfseries 69}
  (2004) 043503} [\href{https://arxiv.org/abs/arXiv:hep-ph/0309033}{{\ttfamily
  arXiv:hep-ph/0309033}}].

\bibitem{Sasaki2006}
M.~{Sasaki}, J.~{V{\"a}liviita} and D.~{Wands}, \emph{{Non-Gaussianity of the
  primordial perturbation in the curvaton model}},
  \href{https://doi.org/10.1103/PhysRevD.74.103003}{\emph{\prd} {\bfseries 74}
  (2006) 103003} [\href{https://arxiv.org/abs/astro-ph/0607627}{{\ttfamily
  astro-ph/0607627}}].

\bibitem{Byrnes2008}
C.~T. {Byrnes}, K.-Y. {Choi} and L.~M.~H. {Hall}, \emph{{Conditions for large
  non-Gaussianity in two-field slow-roll inflation}},
  \href{https://doi.org/10.1088/1475-7516/2008/10/008}{\emph{\jcap} {\bfseries
  10} (2008) 8} [\href{https://arxiv.org/abs/0807.1101}{{\ttfamily
  0807.1101}}].

\bibitem{Byrnes2009a}
C.~T. {Byrnes}, K.-Y. {Choi} and L.~M.~H. {Hall}, \emph{{Large non-Gaussianity
  from two-component hybrid inflation}},
  \href{https://doi.org/10.1088/1475-7516/2009/02/017}{\emph{\jcap} {\bfseries
  2} (2009) 17} [\href{https://arxiv.org/abs/0812.0807}{{\ttfamily
  0812.0807}}].

\bibitem{Byrnes2009b}
C.~T. {Byrnes} and G.~{Tasinato}, \emph{{Non-Gaussianity beyond slow roll in
  multi-field inflation}},
  \href{https://doi.org/10.1088/1475-7516/2009/08/016}{\emph{\jcap} {\bfseries
  8} (2009) 16} [\href{https://arxiv.org/abs/0906.0767}{{\ttfamily
  0906.0767}}].

\bibitem{Byrnes2010}
C.~T. {Byrnes} and K.-Y. {Choi}, \emph{{Review of Local Non-Gaussianity from
  Multifield Inflation}},
  \href{https://doi.org/10.1155/2010/724525}{\emph{Advances in Astronomy}
  {\bfseries 2010} (2010) } [\href{https://arxiv.org/abs/1002.3110}{{\ttfamily
  1002.3110}}].

\bibitem{Linde1996}
A.~{Linde} and V.~{Mukhanov}, \emph{{Non-Gaussian isocurvature perturbations
  from inflation}}, \href{https://doi.org/10.1103/PhysRevD.56.R535}{\emph{\prd}
  {\bfseries 56} (1997) R535}
  [\href{https://arxiv.org/abs/astro-ph/9610219}{{\ttfamily
  astro-ph/9610219}}].

\bibitem{Zaldarriaga2003}
M.~{Zaldarriaga}, \emph{{Non-Gaussianities in models with a varying inflaton
  decay rate}}, \href{https://doi.org/10.1103/PhysRevD.69.043508}{\emph{\prd}
  {\bfseries 69} (2004) 043508}
  [\href{https://arxiv.org/abs/astro-ph/0306006}{{\ttfamily
  astro-ph/0306006}}].

\bibitem{Creminelli2003}
P.~{Creminelli}, \emph{{On non-Gaussianities in single-field inflation}},
  \href{https://doi.org/10.1088/1475-7516/2003/10/003}{\emph{\jcap} {\bfseries
  10} (2003) 003} [\href{https://arxiv.org/abs/astro-ph/0306122}{{\ttfamily
  astro-ph/0306122}}].

\bibitem{Seery2005}
D.~{Seery} and J.~E. {Lidsey}, \emph{{Primordial non-Gaussianities in
  single-field inflation}},
  \href{https://doi.org/10.1088/1475-7516/2005/06/003}{\emph{\jcap} {\bfseries
  6} (2005) 003} [\href{https://arxiv.org/abs/astro-ph/0503692}{{\ttfamily
  astro-ph/0503692}}].

\bibitem{Bartolo2004}
N.~{Bartolo}, E.~{Komatsu}, S.~{Matarrese} and A.~{Riotto},
  \emph{{Non-Gaussianity from inflation: theory and observations}},
  \href{https://doi.org/10.1016/j.physrep.2004.08.022}{\emph{\physrep}
  {\bfseries 402} (2004) 103}
  [\href{https://arxiv.org/abs/astro-ph/0406398}{{\ttfamily
  astro-ph/0406398}}].

\bibitem{Komatsu2009}
E.~{Komatsu}, N.~{Afshordi}, N.~{Bartolo}, D.~{Baumann}, J.~R. {Bond}, E.~I.
  {Buchbinder} et~al., \emph{{Non-Gaussianity as a Probe of the Physics of the
  Primordial Universe and the Astrophysics of the Low Redshift Universe}},
  {\emph{astro2010: The Astronomy and Astrophysics Decadal Survey} {\bfseries
  2010} (2009) 158} [\href{https://arxiv.org/abs/0902.4759}{{\ttfamily
  0902.4759}}].

\bibitem{1905.05697}
{Planck Collaboration}, Y.~{Akrami}, F.~{Arroja}, M.~{Ashdown}, J.~{Aumont},
  C.~{Baccigalupi} et~al., \emph{{Planck 2018 results. IX. Constraints on
  primordial non-Gaussianity}}, {\emph{arXiv e-prints} (2019) arXiv:1905.05697}
  [\href{https://arxiv.org/abs/1905.05697}{{\ttfamily 1905.05697}}].

\bibitem{Slosar2008}
A.~{Slosar}, C.~{Hirata}, U.~{Seljak}, S.~{Ho} and N.~{Padmanabhan},
  \emph{{Constraints on local primordial non-Gaussianity from large scale
  structure}},
  \href{https://doi.org/10.1088/1475-7516/2008/08/031}{\emph{\jcap} {\bfseries
  2008} (2008) 031} [\href{https://arxiv.org/abs/0805.3580}{{\ttfamily
  0805.3580}}].

\bibitem{1208.1491}
A.~J. {Ross}, W.~J. {Percival}, A.~{Carnero}, G.-b. {Zhao}, M.~{Manera},
  A.~{Raccanelli} et~al., \emph{{The clustering of galaxies in the SDSS-III DR9
  Baryon Oscillation Spectroscopic Survey: constraints on primordial
  non-Gaussianity}}, \href{https://doi.org/10.1093/mnras/sts094}{\emph{\mnras}
  {\bfseries 428} (2013) 1116}
  [\href{https://arxiv.org/abs/1208.1491}{{\ttfamily 1208.1491}}].

\bibitem{1303.1349}
T.~{Giannantonio}, A.~J. {Ross}, W.~J. {Percival}, R.~{Crittenden},
  D.~{Bacher}, M.~{Kilbinger} et~al., \emph{{Improved primordial
  non-Gaussianity constraints from measurements of galaxy clustering and the
  integrated Sachs-Wolfe effect}},
  \href{https://doi.org/10.1103/PhysRevD.89.023511}{\emph{\prd} {\bfseries 89}
  (2014) 023511} [\href{https://arxiv.org/abs/1303.1349}{{\ttfamily
  1303.1349}}].

\bibitem{1405.4315}
B.~{Leistedt}, H.~V. {Peiris} and N.~{Roth}, \emph{{Constraints on Primordial
  Non-Gaussianity from 800 000 Photometric Quasars}},
  \href{https://doi.org/10.1103/PhysRevLett.113.221301}{\emph{\prl} {\bfseries
  113} (2014) 221301} [\href{https://arxiv.org/abs/1405.4315}{{\ttfamily
  1405.4315}}].

\bibitem{1310.6716}
D.~{Karagiannis}, T.~{Shanks} and N.~P. {Ross}, \emph{{Search for primordial
  non-Gaussianity in the quasars of SDSS-III BOSS DR9}},
  \href{https://doi.org/10.1093/mnras/stu590}{\emph{\mnras} {\bfseries 441}
  (2014) 486} [\href{https://arxiv.org/abs/1310.6716}{{\ttfamily 1310.6716}}].

\bibitem{1904.08859}
E.~{Castorina}, N.~{Hand}, U.~{Seljak}, F.~{Beutler}, C.-H. {Chuang}, C.~{Zhao}
  et~al., \emph{{Redshift-weighted constraints on primordial non-Gaussianity
  from the clustering of the eBOSS DR14 quasars in Fourier space}},
  \href{https://doi.org/10.1088/1475-7516/2019/09/010}{\emph{\jcap} {\bfseries
  2019} (2019) 010} [\href{https://arxiv.org/abs/1904.08859}{{\ttfamily
  1904.08859}}].

\bibitem{Karagiannis2018}
D.~{Karagiannis}, A.~{Lazanu}, M.~{Liguori}, A.~{Raccanelli}, N.~{Bartolo} and
  L.~{Verde}, \emph{{Constraining primordial non-Gaussianity with bispectrum
  and power spectrum from upcoming optical and radio surveys}},
  \href{https://doi.org/10.1093/mnras/sty1029}{\emph{\mnras} {\bfseries 478}
  (2018) 1341} [\href{https://arxiv.org/abs/1801.09280}{{\ttfamily
  1801.09280}}].

\bibitem{PUMA_surv}
A.~{Slosar}, Z.~{Ahmed}, D.~{Alonso}, M.~A. {Amin}, E.~J. {Arena}, K.~{Bandura}
  et~al., \emph{{Packed Ultra-wideband Mapping Array (PUMA): A Radio Telescope
  for Cosmology and Transients}},  in \emph{\baas}, vol.~51, p.~53, Sep, 2019,
  \href{https://arxiv.org/abs/1907.12559}{{\ttfamily 1907.12559}}.

\bibitem{SKA_redbook}
{\scshape SKA} collaboration, D.~J. Bacon et~al., \emph{{Cosmology with Phase 1
  of the Square Kilometre Array: Red Book 2018: Technical specifications and
  performance forecasts}},
  \href{https://doi.org/10.1017/pasa.2019.51}{\emph{Publ. Astron. Soc.
  Austral.} {\bfseries 37} (2020) e007}
  [\href{https://arxiv.org/abs/1811.02743}{{\ttfamily 1811.02743}}].

\bibitem{Camera2015}
S.~{Camera}, M.~G. {Santos} and R.~{Maartens}, \emph{{Probing primordial
  non-Gaussianity with SKA galaxy redshift surveys: a fully relativistic
  analysis}}, \href{https://doi.org/10.1093/mnras/stv040}{\emph{\mnras}
  {\bfseries 448} (2015) 1035}
  [\href{https://arxiv.org/abs/1409.8286}{{\ttfamily 1409.8286}}].

\bibitem{Camera2013}
S.~{Camera}, M.~G. {Santos}, P.~G. {Ferreira} and L.~{Ferramacho},
  \emph{{Cosmology on Ultralarge Scales with Intensity Mapping of the Neutral
  Hydrogen 21 cm Emission: Limits on Primordial Non-Gaussianity}},
  \href{https://doi.org/10.1103/PhysRevLett.111.171302}{\emph{Physical Review
  Letters} {\bfseries 111} (2013) 171302}
  [\href{https://arxiv.org/abs/1305.6928}{{\ttfamily 1305.6928}}].

\bibitem{Alonso2015}
D.~{Alonso} and P.~G. {Ferreira}, \emph{{Constraining ultralarge-scale
  cosmology with multiple tracers in optical and radio surveys}},
  \href{https://doi.org/10.1103/PhysRevD.92.063525}{\emph{\prd} {\bfseries 92}
  (2015) 063525} [\href{https://arxiv.org/abs/1507.03550}{{\ttfamily
  1507.03550}}].

\bibitem{Dalal2008}
N.~{Dalal}, O.~{Dor{\'e}}, D.~{Huterer} and A.~{Shirokov}, \emph{{Imprints of
  primordial non-Gaussianities on large-scale structure: Scale-dependent bias
  and abundance of virialized objects}},
  \href{https://doi.org/10.1103/PhysRevD.77.123514}{\emph{\prd} {\bfseries 77}
  (2008) 123514} [\href{https://arxiv.org/abs/0710.4560}{{\ttfamily
  0710.4560}}].

\bibitem{Matarrese2008}
S.~{Matarrese} and L.~{Verde}, \emph{{The Effect of Primordial Non-Gaussianity
  on Halo Bias}}, \href{https://doi.org/10.1086/587840}{\emph{\apjl} {\bfseries
  677} (2008) L77} [\href{https://arxiv.org/abs/0801.4826}{{\ttfamily
  0801.4826}}].

\bibitem{Afshordi2008}
N.~{Afshordi} and A.~J. {Tolley}, \emph{{Primordial non-Gaussianity, statistics
  of collapsed objects, and the integrated Sachs-Wolfe effect}},
  \href{https://doi.org/10.1103/PhysRevD.78.123507}{\emph{\prd} {\bfseries 78}
  (2008) 123507} [\href{https://arxiv.org/abs/0806.1046}{{\ttfamily
  0806.1046}}].

\bibitem{Verde2009}
L.~{Verde} and S.~{Matarrese}, \emph{{Detectability of the Effect of
  Inflationary Non-Gaussianity on Halo Bias}},
  \href{https://doi.org/10.1088/0004-637X/706/1/L91}{\emph{\apjl} {\bfseries
  706} (2009) L91} [\href{https://arxiv.org/abs/0909.3224}{{\ttfamily
  0909.3224}}].

\bibitem{Santos2015}
M.~{Santos}, P.~{Bull}, D.~{Alonso}, S.~{Camera}, P.~{Ferreira}, G.~{Bernardi}
  et~al., \emph{{Cosmology from a SKA HI intensity mapping survey}},  in
  \emph{Advancing Astrophysics with the Square Kilometre Array (AASKA14)},
  p.~19, April, 2015, \href{https://arxiv.org/abs/1501.03989}{{\ttfamily
  1501.03989}}.

\bibitem{Cunnington:2020wdu}
S.~Cunnington, S.~Camera and A.~Pourtsidou, \emph{{The degeneracy between
  primordial non-Gaussianity and foregrounds in 21cm intensity mapping
  experiments}},  \href{https://arxiv.org/abs/2007.12126}{{\ttfamily
  2007.12126}}.

\bibitem{Fonseca2015}
J.~{Fonseca}, S.~{Camera}, M.~G. {Santos} and R.~{Maartens}, \emph{{Hunting
  Down Horizon-scale Effects with Multi-wavelength Surveys}},
  \href{https://doi.org/10.1088/2041-8205/812/2/L22}{\emph{\apjl} {\bfseries
  812} (2015) L22} [\href{https://arxiv.org/abs/1507.04605}{{\ttfamily
  1507.04605}}].

\bibitem{Alonso:2015sfa}
D.~Alonso and P.~G. Ferreira, \emph{{Constraining ultralarge-scale cosmology
  with multiple tracers in optical and radio surveys}},
  \href{https://doi.org/10.1103/PhysRevD.92.063525}{\emph{Phys. Rev. D}
  {\bfseries 92} (2015) 063525}
  [\href{https://arxiv.org/abs/1507.03550}{{\ttfamily 1507.03550}}].

\bibitem{Xu2015}
Y.~{Xu}, X.~{Wang} and X.~{Chen}, \emph{{Forecasts on the Dark Energy and
  Primordial Non-Gaussianity Observations with the Tianlai Cylinder Array}},
  \href{https://doi.org/10.1088/0004-637X/798/1/40}{\emph{\apj} {\bfseries 798}
  (2015) 40} [\href{https://arxiv.org/abs/1410.7794}{{\ttfamily 1410.7794}}].

\bibitem{Chen2012}
X.~{Chen}, \emph{{The Tianlai Project: a 21CM Cosmology Experiment}},  in
  \emph{International Journal of Modern Physics Conference Series}, vol.~12 of
  \emph{International Journal of Modern Physics Conference Series},
  pp.~256--263, Mar., 2012, \href{https://arxiv.org/abs/1212.6278}{{\ttfamily
  1212.6278}}, \href{https://doi.org/10.1142/S2010194512006459}{DOI}.

\bibitem{Planck2016_cosmopar}
{Planck Collaboration}, P.~A.~R. {Ade}, N.~{Aghanim}, M.~{Arnaud},
  M.~{Ashdown}, J.~{Aumont} et~al., \emph{{Planck 2015 results. XIII.
  Cosmological parameters}},
  \href{https://doi.org/10.1051/0004-6361/201525830}{\emph{\aap} {\bfseries
  594} (2016) A13} [\href{https://arxiv.org/abs/1502.01589}{{\ttfamily
  1502.01589}}].

\bibitem{CAMB}
A.~{Lewis}, A.~{Challinor} and A.~{Lasenby}, \emph{{Efficient Computation of
  Cosmic Microwave Background Anisotropies in Closed Friedmann-Robertson-Walker
  Models}}, \href{https://doi.org/10.1086/309179}{\emph{\apj} {\bfseries 538}
  (2000) 473} [\href{https://arxiv.org/abs/astro-ph/9911177}{{\ttfamily
  astro-ph/9911177}}].

\bibitem{Salopek1990}
D.~S. {Salopek} and J.~R. {Bond}, \emph{{Nonlinear evolution of long-wavelength
  metric fluctuations in inflationary models}},
  \href{https://doi.org/10.1103/PhysRevD.42.3936}{\emph{\prd} {\bfseries 42}
  (1990) 3936}.

\bibitem{Gangui1993}
A.~{Gangui}, F.~{Lucchin}, S.~{Matarrese} and S.~{Mollerach}, \emph{{The
  three-point correlation function of the cosmic microwave background in
  inflationary models}}, \href{https://doi.org/10.1086/174421}{\emph{\apj}
  {\bfseries 430} (1994) 447}
  [\href{https://arxiv.org/abs/astro-ph/9312033}{{\ttfamily
  astro-ph/9312033}}].

\bibitem{Verde1999}
L.~{Verde}, L.~{Wang}, A.~F. {Heavens} and M.~{Kamionkowski},
  \emph{{Large-scale structure, the cosmic microwave background and primordial
  non-Gaussianity}},
  \href{https://doi.org/10.1046/j.1365-8711.2000.03191.x}{\emph{\mnras}
  {\bfseries 313} (2000) 141}
  [\href{https://arxiv.org/abs/astro-ph/9906301}{{\ttfamily
  astro-ph/9906301}}].

\bibitem{Komatsu2001}
E.~{Komatsu} and D.~N. {Spergel}, \emph{{Acoustic signatures in the primary
  microwave background bispectrum}},
  \href{https://doi.org/10.1103/PhysRevD.63.063002}{\emph{\prd} {\bfseries 63}
  (2001) 063002} [\href{https://arxiv.org/abs/astro-ph/0005036}{{\ttfamily
  astro-ph/0005036}}].

\bibitem{Creminelli2005}
P.~{Creminelli}, A.~{Nicolis}, L.~{Senatore}, M.~{Tegmark} and
  M.~{Zaldarriaga}, \emph{{Limits on non-Gaussianities from WMAP data}},
  \href{https://doi.org/10.1088/1475-7516/2006/05/004}{\emph{\jcap} {\bfseries
  5} (2006) 004} [\href{https://arxiv.org/abs/astro-ph/0509029}{{\ttfamily
  astro-ph/0509029}}].

\bibitem{Senatore2009}
L.~{Senatore}, K.~M. {Smith} and M.~{Zaldarriaga}, \emph{{Non-Gaussianities in
  single field inflation and their optimal limits from the WMAP 5-year data}},
  \href{https://doi.org/10.1088/1475-7516/2010/01/028}{\emph{\jcap} {\bfseries
  1} (2010) 028} [\href{https://arxiv.org/abs/0905.3746}{{\ttfamily
  0905.3746}}].

\bibitem{Bernardeau2002}
F.~{Bernardeau}, S.~{Colombi}, E.~{Gazta{\~n}aga} and R.~{Scoccimarro},
  \emph{{Large-scale structure of the Universe and cosmological perturbation
  theory}},
  \href{https://doi.org/10.1016/S0370-1573(02)00135-7}{\emph{\physrep}
  {\bfseries 367} (2002) 1}
  [\href{https://arxiv.org/abs/astro-ph/0112551}{{\ttfamily
  astro-ph/0112551}}].

\bibitem{Coles1993}
P.~{Coles}, \emph{{Galaxy formation with a local bias}},
  \href{https://doi.org/10.1093/mnras/262.4.1065}{\emph{\mnras} {\bfseries 262}
  (1993) 1065}.

\bibitem{Fry1993}
J.~N. {Fry} and E.~{Gaztanaga}, \emph{{Biasing and hierarchical statistics in
  large-scale structure}}, \href{https://doi.org/10.1086/173015}{\emph{\apj}
  {\bfseries 413} (1993) 447}
  [\href{https://arxiv.org/abs/astro-ph/9302009}{{\ttfamily
  astro-ph/9302009}}].

\bibitem{Fry1996}
J.~N. {Fry}, \emph{{The Evolution of Bias}},
  \href{https://doi.org/10.1086/310006}{\emph{\apjl} {\bfseries 461} (1996)
  L65}.

\bibitem{Catelan1997}
P.~{Catelan}, F.~{Lucchin}, S.~{Matarrese} and C.~{Porciani}, \emph{{The bias
  field of dark matter haloes}},
  \href{https://doi.org/10.1046/j.1365-8711.1998.01455.x}{\emph{\mnras}
  {\bfseries 297} (1998) 692}
  [\href{https://arxiv.org/abs/astro-ph/9708067}{{\ttfamily
  astro-ph/9708067}}].

\bibitem{Catelan2000}
P.~{Catelan}, C.~{Porciani} and M.~{Kamionkowski}, \emph{{Two ways of biasing
  galaxy formation}},
  \href{https://doi.org/10.1046/j.1365-8711.2000.04023.x}{\emph{\mnras}
  {\bfseries 318} (2000) L39}
  [\href{https://arxiv.org/abs/astro-ph/0005544}{{\ttfamily
  astro-ph/0005544}}].

\bibitem{McDonald2009}
P.~{McDonald} and A.~{Roy}, \emph{{Clustering of dark matter tracers:
  generalizing bias for the coming era of precision LSS}},
  \href{https://doi.org/10.1088/1475-7516/2009/08/020}{\emph{\jcap} {\bfseries
  8} (2009) 020} [\href{https://arxiv.org/abs/0902.0991}{{\ttfamily
  0902.0991}}].

\bibitem{Elia2010}
A.~{Elia}, S.~{Kulkarni}, C.~{Porciani}, M.~{Pietroni} and S.~{Matarrese},
  \emph{{Modelling the clustering of dark matter haloes in resummed
  perturbation theories}},
  \href{https://doi.org/10.1111/j.1365-2966.2011.18761.x}{\emph{\mnras}
  {\bfseries 416} (2011) 1703}
  [\href{https://arxiv.org/abs/1012.4833}{{\ttfamily 1012.4833}}].

\bibitem{Chan2012}
K.~C. {Chan}, R.~{Scoccimarro} and R.~K. {Sheth}, \emph{{Gravity and
  large-scale nonlocal bias}},
  \href{https://doi.org/10.1103/PhysRevD.85.083509}{\emph{\prd} {\bfseries 85}
  (2012) 083509} [\href{https://arxiv.org/abs/1201.3614}{{\ttfamily
  1201.3614}}].

\bibitem{Baldauf2012}
T.~{Baldauf}, U.~{Seljak}, V.~{Desjacques} and P.~{McDonald}, \emph{{Evidence
  for quadratic tidal tensor bias from the halo bispectrum}},
  \href{https://doi.org/10.1103/PhysRevD.86.083540}{\emph{\prd} {\bfseries 86}
  (2012) 083540} [\href{https://arxiv.org/abs/1201.4827}{{\ttfamily
  1201.4827}}].

\bibitem{Assassi2014}
V.~{Assassi}, D.~{Baumann}, D.~{Green} and M.~{Zaldarriaga},
  \emph{{Renormalized halo bias}},
  \href{https://doi.org/10.1088/1475-7516/2014/08/056}{\emph{\jcap} {\bfseries
  8} (2014) 056} [\href{https://arxiv.org/abs/1402.5916}{{\ttfamily
  1402.5916}}].

\bibitem{Senatore2014}
L.~{Senatore}, \emph{{Bias in the effective field theory of large scale
  structures}},
  \href{https://doi.org/10.1088/1475-7516/2015/11/007}{\emph{\jcap} {\bfseries
  11} (2015) 007} [\href{https://arxiv.org/abs/1406.7843}{{\ttfamily
  1406.7843}}].

\bibitem{Mirbabayi2014}
M.~{Mirbabayi}, F.~{Schmidt} and M.~{Zaldarriaga}, \emph{{Biased tracers and
  time evolution}},
  \href{https://doi.org/10.1088/1475-7516/2015/07/030}{\emph{\jcap} {\bfseries
  7} (2015) 030} [\href{https://arxiv.org/abs/1412.5169}{{\ttfamily
  1412.5169}}].

\bibitem{Dekel1998}
A.~{Dekel} and O.~{Lahav}, \emph{{Stochastic Nonlinear Galaxy Biasing}},
  \href{https://doi.org/10.1086/307428}{\emph{\apj} {\bfseries 520} (1999) 24}
  [\href{https://arxiv.org/abs/astro-ph/9806193}{{\ttfamily
  astro-ph/9806193}}].

\bibitem{Taruya1998}
A.~{Taruya} and J.~{Soda}, \emph{{Stochastic Biasing and the Galaxy-Mass
  Density Relation in the Weakly Nonlinear Regime}},
  \href{https://doi.org/10.1086/307612}{\emph{\apj} {\bfseries 522} (1999) 46}
  [\href{https://arxiv.org/abs/astro-ph/9809204}{{\ttfamily
  astro-ph/9809204}}].

\bibitem{Matsubara1999}
T.~{Matsubara}, \emph{{Stochasticity of Bias and Nonlocality of Galaxy
  Formation: Linear Scales}}, \href{https://doi.org/10.1086/307931}{\emph{\apj}
  {\bfseries 525} (1999) 543}
  [\href{https://arxiv.org/abs/astro-ph/9906029}{{\ttfamily
  astro-ph/9906029}}].

\bibitem{Desjacques2016}
V.~{Desjacques}, D.~{Jeong} and F.~{Schmidt}, \emph{{Large-scale galaxy bias}},
  \href{https://doi.org/10.1016/j.physrep.2017.12.002}{\emph{\physrep}
  {\bfseries 733} (2018) 1} [\href{https://arxiv.org/abs/1611.09787}{{\ttfamily
  1611.09787}}].

\bibitem{Desjacques:2018pfv}
V.~Desjacques, D.~Jeong and F.~Schmidt, \emph{{The Galaxy Power Spectrum and
  Bispectrum in Redshift Space}},
  \href{https://doi.org/10.1088/1475-7516/2018/12/035}{\emph{JCAP} {\bfseries
  12} (2018) 035} [\href{https://arxiv.org/abs/1806.04015}{{\ttfamily
  1806.04015}}].

\bibitem{Lazeyras2018}
T.~{Lazeyras} and F.~{Schmidt}, \emph{{Beyond LIMD bias: a measurement of the
  complete set of third-order halo bias parameters}},
  \href{https://doi.org/10.1088/1475-7516/2018/09/008}{\emph{\jcap} {\bfseries
  2018} (2018) 008} [\href{https://arxiv.org/abs/1712.07531}{{\ttfamily
  1712.07531}}].

\bibitem{Desjacques2010}
V.~{Desjacques} and U.~{Seljak}, \emph{{Primordial Non-Gaussianity in the
  Large-Scale Structure of the Universe}},
  \href{https://doi.org/10.1155/2010/908640}{\emph{Advances in Astronomy}
  {\bfseries 2010} (2010) 908640}
  [\href{https://arxiv.org/abs/1006.4763}{{\ttfamily 1006.4763}}].

\bibitem{Schmidt2010}
F.~{Schmidt} and M.~{Kamionkowski}, \emph{{Halo clustering with nonlocal
  non-Gaussianity}},
  \href{https://doi.org/10.1103/PhysRevD.82.103002}{\emph{\prd} {\bfseries 82}
  (2010) 103002} [\href{https://arxiv.org/abs/1008.0638}{{\ttfamily
  1008.0638}}].

\bibitem{Scoccimarro2011}
R.~{Scoccimarro}, L.~{Hui}, M.~{Manera} and K.~C. {Chan}, \emph{{Large-scale
  bias and efficient generation of initial conditions for nonlocal primordial
  non-Gaussianity}},
  \href{https://doi.org/10.1103/PhysRevD.85.083002}{\emph{\prd} {\bfseries 85}
  (2012) 083002} [\href{https://arxiv.org/abs/1108.5512}{{\ttfamily
  1108.5512}}].

\bibitem{Desjacques2011b}
V.~{Desjacques}, D.~{Jeong} and F.~{Schmidt}, \emph{{Accurate predictions for
  the scale-dependent galaxy bias from primordial non-Gaussianity}},
  \href{https://doi.org/10.1103/PhysRevD.84.061301}{\emph{\prd} {\bfseries 84}
  (2011) 061301} [\href{https://arxiv.org/abs/1105.3476}{{\ttfamily
  1105.3476}}].

\bibitem{Schmidt2013}
F.~{Schmidt}, D.~{Jeong} and V.~{Desjacques}, \emph{{Peak-background split,
  renormalization, and galaxy clustering}},
  \href{https://doi.org/10.1103/PhysRevD.88.023515}{\emph{\prd} {\bfseries 88}
  (2013) 023515} [\href{https://arxiv.org/abs/1212.0868}{{\ttfamily
  1212.0868}}].

\bibitem{McDonald2008}
P.~{McDonald}, \emph{{Primordial non-Gaussianity: Large-scale structure
  signature in the perturbative bias model}},
  \href{https://doi.org/10.1103/PhysRevD.78.123519}{\emph{\prd} {\bfseries 78}
  (2008) 123519} [\href{https://arxiv.org/abs/0806.1061}{{\ttfamily
  0806.1061}}].

\bibitem{Giannantonio2010}
T.~{Giannantonio} and C.~{Porciani}, \emph{{Structure formation from
  non-Gaussian initial conditions: Multivariate biasing, statistics, and
  comparison with N-body simulations}},
  \href{https://doi.org/10.1103/PhysRevD.81.063530}{\emph{\prd} {\bfseries 81}
  (2010) 063530} [\href{https://arxiv.org/abs/0911.0017}{{\ttfamily
  0911.0017}}].

\bibitem{Baldauf2011}
T.~{Baldauf}, U.~{Seljak} and L.~{Senatore}, \emph{{Primordial non-Gaussianity
  in the bispectrum of the halo density field}},
  \href{https://doi.org/10.1088/1475-7516/2011/04/006}{\emph{\jcap} {\bfseries
  4} (2011) 006} [\href{https://arxiv.org/abs/1011.1513}{{\ttfamily
  1011.1513}}].

\bibitem{Assassi2015}
V.~{Assassi}, D.~{Baumann} and F.~{Schmidt}, \emph{{Galaxy bias and primordial
  non-Gaussianity}},
  \href{https://doi.org/10.1088/1475-7516/2015/12/043}{\emph{\jcap} {\bfseries
  12} (2015) 043} [\href{https://arxiv.org/abs/1510.03723}{{\ttfamily
  1510.03723}}].

\bibitem{PS1973}
W.~H. {Press} and P.~{Schechter}, \emph{{Formation of Galaxies and Clusters of
  Galaxies by Self-Similar Gravitational Condensation}},
  \href{https://doi.org/10.1086/152650}{\emph{\apj} {\bfseries 187} (1974)
  425}.

\bibitem{Tinker2008}
J.~{Tinker}, A.~V. {Kravtsov}, A.~{Klypin}, K.~{Abazajian}, M.~{Warren},
  G.~{Yepes} et~al., \emph{{Toward a Halo Mass Function for Precision
  Cosmology: The Limits of Universality}},
  \href{https://doi.org/10.1086/591439}{\emph{\apj} {\bfseries 688} (2008) 709}
  [\href{https://arxiv.org/abs/0803.2706}{{\ttfamily 0803.2706}}].

\bibitem{Desjacques2011a}
V.~{Desjacques}, D.~{Jeong} and F.~{Schmidt}, \emph{{Non-Gaussian Halo Bias
  Re-examined: Mass-dependent Amplitude from the Peak-Background Split and
  Thresholding}}, \href{https://doi.org/10.1103/PhysRevD.84.063512}{\emph{\prd}
  {\bfseries 84} (2011) 063512}
  [\href{https://arxiv.org/abs/1105.3628}{{\ttfamily 1105.3628}}].

\bibitem{Sefusatti2012}
E.~{Sefusatti}, M.~{Crocce} and V.~{Desjacques}, \emph{{The halo bispectrum in
  N-body simulations with non-Gaussian initial conditions}},
  \href{https://doi.org/10.1111/j.1365-2966.2012.21271.x}{\emph{\mnras}
  {\bfseries 425} (2012) 2903}
  [\href{https://arxiv.org/abs/1111.6966}{{\ttfamily 1111.6966}}].

\bibitem{Giannantonio2012}
T.~{Giannantonio}, C.~{Porciani}, J.~{Carron}, A.~{Amara} and A.~{Pillepich},
  \emph{{Constraining primordial non-Gaussianity with future galaxy surveys}},
  \href{https://doi.org/10.1111/j.1365-2966.2012.20604.x}{\emph{\mnras}
  {\bfseries 422} (2012) 2854}
  [\href{https://arxiv.org/abs/1109.0958}{{\ttfamily 1109.0958}}].

\bibitem{LoVerde2007}
M.~{LoVerde}, A.~{Miller}, S.~{Shandera} and L.~{Verde}, \emph{{Effects of
  scale-dependent non-Gaussianity on cosmological structures}},
  \href{https://doi.org/10.1088/1475-7516/2008/04/014}{\emph{\jcap} {\bfseries
  4} (2008) 014} [\href{https://arxiv.org/abs/0711.4126}{{\ttfamily
  0711.4126}}].

\bibitem{Desjacques2009}
V.~{Desjacques}, U.~{Seljak} and I.~T. {Iliev}, \emph{{Scale-dependent bias
  induced by local non-Gaussianity: a comparison to N-body simulations}},
  \href{https://doi.org/10.1111/j.1365-2966.2009.14721.x}{\emph{\mnras}
  {\bfseries 396} (2009) 85} [\href{https://arxiv.org/abs/0811.2748}{{\ttfamily
  0811.2748}}].

\bibitem{Seljak2000}
U.~Seljak, \emph{{Analytic model for galaxy and dark matter clustering}},
  \href{https://doi.org/10.1046/j.1365-8711.2000.03715.x}{\emph{Mon. Not. Roy.
  Astron. Soc.} {\bfseries 318} (2000) 203}
  [\href{https://arxiv.org/abs/astro-ph/0001493}{{\ttfamily
  astro-ph/0001493}}].

\bibitem{Peacock2000}
J.~Peacock and R.~Smith, \emph{{Halo occupation numbers and galaxy bias}},
  \href{https://doi.org/10.1046/j.1365-8711.2000.03779.x}{\emph{Mon. Not. Roy.
  Astron. Soc.} {\bfseries 318} (2000) 1144}
  [\href{https://arxiv.org/abs/astro-ph/0005010}{{\ttfamily
  astro-ph/0005010}}].

\bibitem{Scoccimarro2000}
R.~Scoccimarro, R.~K. Sheth, L.~Hui and B.~Jain, \emph{{How many galaxies fit
  in a halo? Constraints on galaxy formation efficiency from spatial
  clustering}}, \href{https://doi.org/10.1086/318261}{\emph{Astrophys. J.}
  {\bfseries 546} (2001) 20}
  [\href{https://arxiv.org/abs/astro-ph/0006319}{{\ttfamily
  astro-ph/0006319}}].

\bibitem{Villaescusa2014}
F.~Villaescusa-Navarro, M.~Viel, K.~K. Datta and T.~R. Choudhury,
  \emph{{Modeling the neutral hydrogen distribution in the post-reionization
  Universe: intensity mapping}},
  \href{https://doi.org/10.1088/1475-7516/2014/09/050}{\emph{JCAP} {\bfseries
  09} (2014) 050} [\href{https://arxiv.org/abs/1405.6713}{{\ttfamily
  1405.6713}}].

\bibitem{Castorina2016}
E.~{Castorina} and F.~{Villaescusa-Navarro}, \emph{{On the spatial distribution
  of neutral hydrogen in the Universe: bias and shot-noise of the H I power
  spectrum}}, \href{https://doi.org/10.1093/mnras/stx1599}{\emph{\mnras}
  {\bfseries 471} (2017) 1788}
  [\href{https://arxiv.org/abs/1609.05157}{{\ttfamily 1609.05157}}].

\bibitem{Tinker2010}
J.~L. {Tinker}, B.~E. {Robertson}, A.~V. {Kravtsov}, A.~{Klypin}, M.~S.
  {Warren}, G.~{Yepes} et~al., \emph{{The Large-scale Bias of Dark Matter
  Halos: Numerical Calibration and Model Tests}},
  \href{https://doi.org/10.1088/0004-637X/724/2/878}{\emph{\apj} {\bfseries
  724} (2010) 878} [\href{https://arxiv.org/abs/1001.3162}{{\ttfamily
  1001.3162}}].

\bibitem{Lazeyras2016}
T.~{Lazeyras}, C.~{Wagner}, T.~{Baldauf} and F.~{Schmidt}, \emph{{Precision
  measurement of the local bias of dark matter halos}},
  \href{https://doi.org/10.1088/1475-7516/2016/02/018}{\emph{\jcap} {\bfseries
  2016} (2016) 018} [\href{https://arxiv.org/abs/1511.01096}{{\ttfamily
  1511.01096}}].

\bibitem{ShethTormen1999}
R.~K. {Sheth} and G.~{Tormen}, \emph{{Large-scale bias and the peak background
  split}},
  \href{https://doi.org/10.1046/j.1365-8711.1999.02692.x}{\emph{\mnras}
  {\bfseries 308} (1999) 119}
  [\href{https://arxiv.org/abs/astro-ph/9901122}{{\ttfamily
  astro-ph/9901122}}].

\bibitem{Padmanabhan2017a}
H.~{Padmanabhan} and A.~{Refregier}, \emph{{Constraining a halo model for
  cosmological neutral hydrogen}},
  \href{https://doi.org/10.1093/mnras/stw2706}{\emph{\mnras} {\bfseries 464}
  (2017) 4008} [\href{https://arxiv.org/abs/1607.01021}{{\ttfamily
  1607.01021}}].

\bibitem{Padmanabhan2017b}
H.~{Padmanabhan}, A.~{Refregier} and A.~{Amara}, \emph{{A halo model for
  cosmological neutral hydrogen : abundances and clustering}},
  \href{https://doi.org/10.1093/mnras/stx979}{\emph{\mnras} {\bfseries 469}
  (2017) 2323} [\href{https://arxiv.org/abs/1611.06235}{{\ttfamily
  1611.06235}}].

\bibitem{Sargent:1977}
W.~L.~W. {Sargent} and E.~L. {Turner}, \emph{{A statistical method for
  determining the cosmological density parameter from the redshifts of a
  complete sample of galaxies}},
  \href{https://doi.org/10.1086/182362}{\emph{Astrophys. J.} {\bfseries 212}
  (1977) L3}.

\bibitem{Kaiser1987}
N.~{Kaiser}, \emph{{Clustering in real space and in redshift space}},
  \href{https://doi.org/10.1093/mnras/227.1.1}{\emph{\mnras} {\bfseries 227}
  (1987) 1}.

\bibitem{Hamilton1998}
A.~J.~S. {Hamilton}, \emph{{Linear Redshift Distortions: a Review}},  in
  \emph{The Evolving Universe} (D.~{Hamilton}, ed.), vol.~231 of
  \emph{Astrophysics and Space Science Library}, p.~185, 1998,
  \href{https://arxiv.org/abs/astro-ph/9708102}{{\ttfamily astro-ph/9708102}},
  \href{https://doi.org/10.1007/978-94-011-4960-0_17}{DOI}.

\bibitem{Verde1998}
L.~{Verde}, A.~F. {Heavens}, S.~{Matarrese} and L.~{Moscardini},
  \emph{{Large-scale bias in the Universe - II. Redshift-space bispectrum}},
  \href{https://doi.org/10.1046/j.1365-8711.1998.01937.x}{\emph{\mnras}
  {\bfseries 300} (1998) 747}
  [\href{https://arxiv.org/abs/astro-ph/9806028}{{\ttfamily
  astro-ph/9806028}}].

\bibitem{Scoccimarro1998}
R.~{Scoccimarro}, S.~{Colombi}, J.~N. {Fry}, J.~A. {Frieman}, E.~{Hivon} and
  A.~{Melott}, \emph{{Nonlinear Evolution of the Bispectrum of Cosmological
  Perturbations}}, \href{https://doi.org/10.1086/305399}{\emph{\apj} {\bfseries
  496} (1998) 586} [\href{https://arxiv.org/abs/astro-ph/9704075}{{\ttfamily
  astro-ph/9704075}}].

\bibitem{Jackson1972}
J.~C. {Jackson}, \emph{{A critique of Rees's theory of primordial gravitational
  radiation}}, \href{https://doi.org/10.1093/mnras/156.1.1P}{\emph{\mnras}
  {\bfseries 156} (1972) 1P} [\href{https://arxiv.org/abs/0810.3908}{{\ttfamily
  0810.3908}}].

\bibitem{Gil-Marin:2014sta}
H.~Gil-Marín, J.~Noreña, L.~Verde, W.~J. Percival, C.~Wagner, M.~Manera
  et~al., \emph{{The power spectrum and bispectrum of SDSS DR11 BOSS galaxies
  -- I. Bias and gravity}},
  \href{https://doi.org/10.1093/mnras/stv961}{\emph{Mon. Not. Roy. Astron.
  Soc.} {\bfseries 451} (2015) 539}
  [\href{https://arxiv.org/abs/1407.5668}{{\ttfamily 1407.5668}}].

\bibitem{Hashimoto:2017klo}
I.~Hashimoto, Y.~Rasera and A.~Taruya, \emph{{Precision cosmology with
  redshift-space bispectrum: a perturbation theory based model at one-loop
  order}}, \href{https://doi.org/10.1103/PhysRevD.96.043526}{\emph{Phys. Rev.
  D} {\bfseries 96} (2017) 043526}
  [\href{https://arxiv.org/abs/1705.02574}{{\ttfamily 1705.02574}}].

\bibitem{Lazanu:2018yae}
A.~Lazanu and M.~Liguori, \emph{{The two and three-loop matter bispectrum in
  perturbation theories}},
  \href{https://doi.org/10.1088/1475-7516/2018/04/055}{\emph{JCAP} {\bfseries
  04} (2018) 055} [\href{https://arxiv.org/abs/1803.03184}{{\ttfamily
  1803.03184}}].

\bibitem{Oddo:2019run}
A.~Oddo, E.~Sefusatti, C.~Porciani, P.~Monaco and A.~G. Sánchez, \emph{{Toward
  a robust inference method for the galaxy bispectrum: likelihood function and
  model selection}},
  \href{https://doi.org/10.1088/1475-7516/2020/03/056}{\emph{JCAP} {\bfseries
  03} (2020) 056} [\href{https://arxiv.org/abs/1908.01774}{{\ttfamily
  1908.01774}}].

\bibitem{Scoccimarro1999}
R.~{Scoccimarro}, H.~M.~P. {Couchman} and J.~A. {Frieman}, \emph{{The
  Bispectrum as a Signature of Gravitational Instability in Redshift Space}},
  \href{https://doi.org/10.1086/307220}{\emph{\apj} {\bfseries 517} (1999) 531}
  [\href{https://arxiv.org/abs/astro-ph/9808305}{{\ttfamily
  astro-ph/9808305}}].

\bibitem{StageII2018}
{Cosmic Visions 21 cm Collaboration}, R.~{Ansari}, E.~J. {Arena}, K.~{Bandura},
  P.~{Bull}, E.~{Castorina} et~al., \emph{{Inflation and Early Dark Energy with
  a Stage II Hydrogen Intensity Mapping Experiment}}, {\emph{arXiv e-prints}
  (2018) arXiv:1810.09572} [\href{https://arxiv.org/abs/1810.09572}{{\ttfamily
  1810.09572}}].

\bibitem{Schmidt2015}
F.~{Schmidt}, \emph{{Towards a self-consistent halo model for the nonlinear
  large-scale structure}},
  \href{https://doi.org/10.1103/PhysRevD.93.063512}{\emph{\prd} {\bfseries 93}
  (2016) 063512} [\href{https://arxiv.org/abs/1511.02231}{{\ttfamily
  1511.02231}}].

\bibitem{Bertacca2017}
D.~{Bertacca}, A.~{Raccanelli}, N.~{Bartolo}, M.~{Liguori}, S.~{Matarrese} and
  L.~{Verde}, \emph{{Relativistic wide-angle galaxy bispectrum on the light
  cone}}, \href{https://doi.org/10.1103/PhysRevD.97.023531}{\emph{\prd}
  {\bfseries 97} (2018) 023531}
  [\href{https://arxiv.org/abs/1705.09306}{{\ttfamily 1705.09306}}].

\bibitem{Zaldarriaga2003b}
M.~{Zaldarriaga}, S.~R. {Furlanetto} and L.~{Hernquist}, \emph{{21 Centimeter
  Fluctuations from Cosmic Gas at High Redshifts}},
  \href{https://doi.org/10.1086/386327}{\emph{\apj} {\bfseries 608} (2004) 622}
  [\href{https://arxiv.org/abs/astro-ph/0311514}{{\ttfamily
  astro-ph/0311514}}].

\bibitem{Tegmark2008}
M.~{Tegmark} and M.~{Zaldarriaga}, \emph{{Fast Fourier transform telescope}},
  \href{https://doi.org/10.1103/PhysRevD.79.083530}{\emph{\prd} {\bfseries 79}
  (2009) 083530} [\href{https://arxiv.org/abs/0805.4414}{{\ttfamily
  0805.4414}}].

\bibitem{Newburgh2014}
L.~B. {Newburgh}, G.~E. {Addison}, M.~{Amiri}, K.~{Bandura}, J.~R. {Bond},
  L.~{Connor} et~al., \emph{{Calibrating CHIME: a new radio interferometer to
  probe dark energy}},  in \emph{\procspie}, vol.~9145 of \emph{Society of
  Photo-Optical Instrumentation Engineers (SPIE) Conference Series}, p.~91454V,
  Jul, 2014, \href{https://arxiv.org/abs/1406.2267}{{\ttfamily 1406.2267}},
  \href{https://doi.org/10.1117/12.2056962}{DOI}.

\bibitem{Newburgh2016}
L.~B. {Newburgh}, K.~{Bandura}, M.~A. {Bucher}, T.~C. {Chang}, H.~C. {Chiang},
  J.~F. {Cliche} et~al., \emph{{HIRAX: a probe of dark energy and radio
  transients}},  in \emph{\procspie}, vol.~9906 of \emph{Society of
  Photo-Optical Instrumentation Engineers (SPIE) Conference Series}, p.~99065X,
  Aug, 2016, \href{https://arxiv.org/abs/1607.02059}{{\ttfamily 1607.02059}},
  \href{https://doi.org/10.1117/12.2234286}{DOI}.

\bibitem{Pourtsidou:2016ctq}
A.~Pourtsidou, \emph{{Synergistic tests of inflation}},
  \href{https://arxiv.org/abs/1612.05138}{{\ttfamily 1612.05138}}.

\bibitem{Shaw:2013wza}
J.~Shaw, K.~Sigurdson, U.-L. Pen, A.~Stebbins and M.~Sitwell, \emph{{All-Sky
  Interferometry with Spherical Harmonic Transit Telescopes}},
  \href{https://doi.org/10.1088/0004-637X/781/2/57}{\emph{Astrophys. J.}
  {\bfseries 781} (2014) 57} [\href{https://arxiv.org/abs/1302.0327}{{\ttfamily
  1302.0327}}].

\bibitem{Shaw:2014khi}
J.~Shaw, K.~Sigurdson, M.~Sitwell, A.~Stebbins and U.-L. Pen, \emph{{Coaxing
  cosmic 21 cm fluctuations from the polarized sky using m-mode analysis}},
  \href{https://doi.org/10.1103/PhysRevD.91.083514}{\emph{Phys. Rev. D}
  {\bfseries 91} (2015) 083514}
  [\href{https://arxiv.org/abs/1401.2095}{{\ttfamily 1401.2095}}].

\bibitem{Pober:2014lva}
J.~C. Pober, \emph{{The Impact of Foregrounds on Redshift Space Distortion
  Measurements With the Highly-Redshifted 21 cm Line}},
  \href{https://doi.org/10.1093/mnras/stu2575}{\emph{Mon. Not. Roy. Astron.
  Soc.} {\bfseries 447} (2015) 1705}
  [\href{https://arxiv.org/abs/1411.2050}{{\ttfamily 1411.2050}}].

\bibitem{Byrne:2018dkh}
R.~Byrne, M.~F. Morales, B.~Hazelton, W.~Li, N.~Barry, A.~P. Beardsley et~al.,
  \emph{{Fundamental Limitations on the Calibration of Redundant 21 cm
  Cosmology Instruments and Implications for HERA and the SKA}},
  \href{https://doi.org/10.3847/1538-4357/ab107d}{\emph{Astrophys. J.}
  {\bfseries 875} (2019) 70}
  [\href{https://arxiv.org/abs/1811.01378}{{\ttfamily 1811.01378}}].

\bibitem{Jacobson:2003wv}
T.~Jacobson and R.~Parentani, \emph{{Horizon entropy}},
  \href{https://doi.org/10.1023/A:1023785123428}{\emph{Found. Phys.} {\bfseries
  33} (2003) 323} [\href{https://arxiv.org/abs/gr-qc/0302099}{{\ttfamily
  gr-qc/0302099}}].

\bibitem{Furlanetto:2006jb}
S.~Furlanetto, S.~Oh and F.~Briggs, \emph{{Cosmology at Low Frequencies: The 21
  cm Transition and the High-Redshift Universe}},
  \href{https://doi.org/10.1016/j.physrep.2006.08.002}{\emph{Phys. Rept.}
  {\bfseries 433} (2006) 181}
  [\href{https://arxiv.org/abs/astro-ph/0608032}{{\ttfamily
  astro-ph/0608032}}].

\bibitem{Liu2011}
A.~{Liu} and M.~{Tegmark}, \emph{{A method for 21 cm power spectrum estimation
  in the presence of foregrounds}},
  \href{https://doi.org/10.1103/PhysRevD.83.103006}{\emph{\prd} {\bfseries 83}
  (2011) 103006} [\href{https://arxiv.org/abs/1103.0281}{{\ttfamily
  1103.0281}}].

\bibitem{Liu2012}
A.~{Liu} and M.~{Tegmark}, \emph{{How well can we measure and understand
  foregrounds with 21-cm experiments?}},
  \href{https://doi.org/10.1111/j.1365-2966.2011.19989.x}{\emph{\mnras}
  {\bfseries 419} (2012) 3491}
  [\href{https://arxiv.org/abs/1106.0007}{{\ttfamily 1106.0007}}].

\bibitem{Jasche2010}
J.~{Jasche} and F.~S. {Kitaura}, \emph{{Fast Hamiltonian sampling for
  large-scale structure inference}},
  \href{https://doi.org/10.1111/j.1365-2966.2010.16897.x}{\emph{\mnras}
  {\bfseries 407} (2010) 29} [\href{https://arxiv.org/abs/0911.2496}{{\ttfamily
  0911.2496}}].

\bibitem{Kitaura2013}
F.~S. {Kitaura}, \emph{{The initial conditions of the universe from constrained
  simulations.}}, \href{https://doi.org/10.1093/mnrasl/sls029}{\emph{\mnras}
  {\bfseries 429} (2013) L84}
  [\href{https://arxiv.org/abs/1203.4184}{{\ttfamily 1203.4184}}].

\bibitem{Wang:2014hia}
H.~Wang, H.~Mo, X.~Yang, Y.~Jing and W.~Lin, \emph{{ELUCID - Exploring the
  Local Universe with reConstructed Initial Density field I: Hamiltonian Markov
  Chain Monte Carlo Method with Particle Mesh Dynamics}},
  \href{https://doi.org/10.1088/0004-637X/794/1/94}{\emph{Astrophys. J.}
  {\bfseries 794} (2014) 94} [\href{https://arxiv.org/abs/1407.3451}{{\ttfamily
  1407.3451}}].

\bibitem{Jasche:2014vpa}
J.~Jasche, F.~Leclercq and B.~D. Wandelt, \emph{{Past and present cosmic
  structure in the SDSS DR7 main sample}},
  \href{https://doi.org/10.1088/1475-7516/2015/01/036}{\emph{JCAP} {\bfseries
  01} (2015) 036} [\href{https://arxiv.org/abs/1409.6308}{{\ttfamily
  1409.6308}}].

\bibitem{Wang:2016qbz}
H.~Wang, H.~Mo, X.~Yang, Y.~Zhang, J.~Shi, Y.~Jing et~al., \emph{{ELUCID -
  Exploring the Local Universe with reConstructed Initial Density field III:
  Constrained Simulation in the SDSS Volume}},
  \href{https://doi.org/10.3847/0004-637X/831/2/164}{\emph{Astrophys. J.}
  {\bfseries 831} (2016) 164}
  [\href{https://arxiv.org/abs/1608.01763}{{\ttfamily 1608.01763}}].

\bibitem{Seljak:2017rmr}
U.~Seljak, G.~Aslanyan, Y.~Feng and C.~Modi, \emph{{Towards optimal extraction
  of cosmological information from nonlinear data}},
  \href{https://doi.org/10.1088/1475-7516/2017/12/009}{\emph{JCAP} {\bfseries
  12} (2017) 009} [\href{https://arxiv.org/abs/1706.06645}{{\ttfamily
  1706.06645}}].

\bibitem{Modi:2018cfi}
C.~Modi, Y.~Feng and U.~Seljak, \emph{{Cosmological Reconstruction From Galaxy
  Light: Neural Network Based Light-Matter Connection}},
  \href{https://doi.org/10.1088/1475-7516/2018/10/028}{\emph{JCAP} {\bfseries
  10} (2018) 028} [\href{https://arxiv.org/abs/1805.02247}{{\ttfamily
  1805.02247}}].

\bibitem{Zhu:2016esh}
H.-M. Zhu, U.-L. Pen, Y.~Yu and X.~Chen, \emph{{Recovering lost 21 cm radial
  modes via cosmic tidal reconstruction}},
  \href{https://doi.org/10.1103/PhysRevD.98.043511}{\emph{Phys. Rev. D}
  {\bfseries 98} (2018) 043511}
  [\href{https://arxiv.org/abs/1610.07062}{{\ttfamily 1610.07062}}].

\bibitem{Karacayli:2019iyd}
N.~G. Karaçayl\i and N.~Padmanabhan, \emph{{Anatomy of Cosmic Tidal
  Reconstruction}}, \href{https://doi.org/10.1093/mnras/stz964}{\emph{Mon. Not.
  Roy. Astron. Soc.} {\bfseries 486} (2019) 3864}
  [\href{https://arxiv.org/abs/1904.01387}{{\ttfamily 1904.01387}}].

\bibitem{Modi:2019hnu}
C.~Modi, M.~White, A.~Slosar and E.~Castorina, \emph{{Reconstructing
  large-scale structure with neutral hydrogen surveys}},
  \href{https://doi.org/10.1088/1475-7516/2019/11/023}{\emph{JCAP} {\bfseries
  11} (2019) 023} [\href{https://arxiv.org/abs/1907.02330}{{\ttfamily
  1907.02330}}].

\bibitem{Pober2014}
J.~C. {Pober}, A.~{Liu}, J.~S. {Dillon}, J.~E. {Aguirre}, J.~D. {Bowman}, R.~F.
  {Bradley} et~al., \emph{{What Next-generation 21 cm Power Spectrum
  Measurements can Teach us About the Epoch of Reionization}},
  \href{https://doi.org/10.1088/0004-637X/782/2/66}{\emph{\apj} {\bfseries 782}
  (2014) 66} [\href{https://arxiv.org/abs/1310.7031}{{\ttfamily 1310.7031}}].

\bibitem{Pober2015}
J.~C. {Pober}, \emph{{The impact of foregrounds on redshift space distortion
  measurements with the highly redshifted 21-cm line}},
  \href{https://doi.org/10.1093/mnras/stu2575}{\emph{\mnras} {\bfseries 447}
  (2015) 1705} [\href{https://arxiv.org/abs/1411.2050}{{\ttfamily 1411.2050}}].

\bibitem{Scoccimarro2003}
R.~{Scoccimarro}, E.~{Sefusatti} and M.~{Zaldarriaga}, \emph{{Probing
  primordial non-Gaussianity with large-scale structure}},
  \href{https://doi.org/10.1103/PhysRevD.69.103513}{\emph{\prd} {\bfseries 69}
  (2004) 103513} [\href{https://arxiv.org/abs/astro-ph/0312286}{{\ttfamily
  astro-ph/0312286}}].

\bibitem{Sefusatti2006}
E.~{Sefusatti}, M.~{Crocce}, S.~{Pueblas} and R.~{Scoccimarro},
  \emph{{Cosmology and the bispectrum}},
  \href{https://doi.org/10.1103/PhysRevD.74.023522}{\emph{\prd} {\bfseries 74}
  (2006) 023522} [\href{https://arxiv.org/abs/astro-ph/0604505}{{\ttfamily
  astro-ph/0604505}}].

\bibitem{Sefusatti2007}
E.~{Sefusatti} and E.~{Komatsu}, \emph{{Bispectrum of galaxies from
  high-redshift galaxy surveys: Primordial non-Gaussianity and nonlinear galaxy
  bias}}, \href{https://doi.org/10.1103/PhysRevD.76.083004}{\emph{\prd}
  {\bfseries 76} (2007) 083004}
  [\href{https://arxiv.org/abs/0705.0343}{{\ttfamily 0705.0343}}].

\bibitem{Chan2017}
K.~C. {Chan} and L.~{Blot}, \emph{{Assessment of the information content of the
  power spectrum and bispectrum}},
  \href{https://doi.org/10.1103/PhysRevD.96.023528}{\emph{\prd} {\bfseries 96}
  (2017) 023528} [\href{https://arxiv.org/abs/1610.06585}{{\ttfamily
  1610.06585}}].

\bibitem{Smith2003}
R.~E. {Smith}, J.~A. {Peacock}, A.~{Jenkins}, S.~D.~M. {White}, C.~S. {Frenk},
  F.~R. {Pearce} et~al., \emph{{Stable clustering, the halo model and
  non-linear cosmological power spectra}},
  \href{https://doi.org/10.1046/j.1365-8711.2003.06503.x}{\emph{\mnras}
  {\bfseries 341} (2003) 1311}
  [\href{https://arxiv.org/abs/astro-ph/0207664}{{\ttfamily
  astro-ph/0207664}}].

\bibitem{Takahashi2012}
R.~{Takahashi}, M.~{Sato}, T.~{Nishimichi}, A.~{Taruya} and M.~{Oguri},
  \emph{{Revising the Halofit Model for the Nonlinear Matter Power Spectrum}},
  \href{https://doi.org/10.1088/0004-637X/761/2/152}{\emph{\apj} {\bfseries
  761} (2012) 152} [\href{https://arxiv.org/abs/1208.2701}{{\ttfamily
  1208.2701}}].

\bibitem{Wang2006}
Y.~{Wang}, \emph{{Dark Energy Constraints from Baryon Acoustic Oscillations}},
  \href{https://doi.org/10.1086/505384}{\emph{\apj} {\bfseries 647} (2006) 1}
  [\href{https://arxiv.org/abs/astro-ph/0601163}{{\ttfamily
  astro-ph/0601163}}].

\bibitem{Liguori2008}
M.~{Liguori} and A.~{Riotto}, \emph{{Impact of uncertainties in the
  cosmological parameters on the measurement of primordial non-Gaussianity}},
  \href{https://doi.org/10.1103/PhysRevD.78.123004}{\emph{\prd} {\bfseries 78}
  (2008) 123004} [\href{https://arxiv.org/abs/0808.3255}{{\ttfamily
  0808.3255}}].

\bibitem{Moradinezhad2018}
A.~{Moradinezhad Dizgah}, H.~{Lee}, J.~B. {Mu{\~n}oz} and C.~{Dvorkin},
  \emph{{Galaxy Bispectrum from Massive Spinning Particles}}, {\emph{ArXiv
  e-prints} (2018) } [\href{https://arxiv.org/abs/1801.07265}{{\ttfamily
  1801.07265}}].

\bibitem{Bellomo:2020pnw}
N.~Bellomo, J.~L. Bernal, G.~Scelfo, A.~Raccanelli and L.~Verde, \emph{{Beware
  of commonly used approximations I: errors in forecasts}},
  \href{https://arxiv.org/abs/2005.10384}{{\ttfamily 2005.10384}}.

\bibitem{Agarwal:2020lov}
N.~Agarwal, V.~Desjacques, D.~Jeong and F.~Schmidt, \emph{{Information content
  in the redshift-space galaxy power spectrum and bispectrum}},
  \href{https://arxiv.org/abs/2007.04340}{{\ttfamily 2007.04340}}.

\bibitem{Baldauf2016}
T.~{Baldauf}, M.~{Mirbabayi}, M.~{Simonovi{\'c}} and M.~{Zaldarriaga},
  \emph{{LSS constraints with controlled theoretical uncertainties}},
  {\emph{ArXiv e-prints} (2016) }
  [\href{https://arxiv.org/abs/1602.00674}{{\ttfamily 1602.00674}}].

\bibitem{Bull2015}
P.~{Bull}, P.~G. {Ferreira}, P.~{Patel} and M.~G. {Santos}, \emph{{Late-time
  Cosmology with 21 cm Intensity Mapping Experiments}},
  \href{https://doi.org/10.1088/0004-637X/803/1/21}{\emph{\apj} {\bfseries 803}
  (2015) 21} [\href{https://arxiv.org/abs/1405.1452}{{\ttfamily 1405.1452}}].

\bibitem{Watkinson:2020zqg}
C.~A. Watkinson, C.~M. Trott and I.~Hothi, \emph{{The bispectrum and 21cm
  foregrounds during the Epoch of Reionization}},
  \href{https://arxiv.org/abs/2002.05992}{{\ttfamily 2002.05992}}.

\bibitem{Planck_PNG2016}
{Planck Collaboration}, P.~A.~R. {Ade}, N.~{Aghanim}, M.~{Arnaud}, F.~{Arroja},
  M.~{Ashdown} et~al., \emph{{Planck 2015 results. XVII. Constraints on
  primordial non-Gaussianity}},
  \href{https://doi.org/10.1051/0004-6361/201525836}{\emph{\aap} {\bfseries
  594} (2016) A17} [\href{https://arxiv.org/abs/1502.01592}{{\ttfamily
  1502.01592}}].

\bibitem{Dore2014}
O.~{Dor{\'e}}, J.~{Bock}, M.~{Ashby}, P.~{Capak}, A.~{Cooray}, R.~{de Putter}
  et~al., \emph{{Cosmology with the SPHEREX All-Sky Spectral Survey}},
  {\emph{ArXiv e-prints} (2014) }
  [\href{https://arxiv.org/abs/1412.4872}{{\ttfamily 1412.4872}}].

\bibitem{Tellarini2016}
M.~{Tellarini}, A.~J. {Ross}, G.~{Tasinato} and D.~{Wands}, \emph{{Galaxy
  bispectrum, primordial non-Gaussianity and redshift space distortions}},
  \href{https://doi.org/10.1088/1475-7516/2016/06/014}{\emph{\jcap} {\bfseries
  6} (2016) 014} [\href{https://arxiv.org/abs/1603.06814}{{\ttfamily
  1603.06814}}].

\bibitem{Peacock1994}
J.~A. {Peacock} and S.~J. {Dodds}, \emph{{Reconstructing the Linear Power
  Spectrum of Cosmological Mass Fluctuations}},
  \href{https://doi.org/10.1093/mnras/267.4.1020}{\emph{\mnras} {\bfseries 267}
  (1994) 1020} [\href{https://arxiv.org/abs/astro-ph/9311057}{{\ttfamily
  astro-ph/9311057}}].

\bibitem{Ballinger1996}
W.~E. {Ballinger}, J.~A. {Peacock} and A.~F. {Heavens}, \emph{{Measuring the
  cosmological constant with redshift surveys}},
  \href{https://doi.org/10.1093/mnras/282.3.877}{\emph{\mnras} {\bfseries 282}
  (1996) 877} [\href{https://arxiv.org/abs/astro-ph/9605017}{{\ttfamily
  astro-ph/9605017}}].

\end{thebibliography}\endgroup

\appendix

\section{Derivation of Bias Parameters from the Tinker et al mass function with the PBS approach}\label{app:PBSbias}

The peak-background split (PBS) approach (see \eg Ref. \cite{Desjacques2016} for a review) derives the halo bias parameters from the change in the distribution of the density peaks (\ie the mass function). The density fluctuation field can be decomposed into a long-wavelength linear fluctuation , $\delta_l(\mathbf{x})$, and a noisy short wavelength one, $\delta_s(\mathbf{x})$. The first will modulate the background density and alter the height of the peaks to an effective value

\begin{equation}\label{eq:neff}
   \nu\rightarrow\nu_{\text{eff}}=\frac{\delta_c-\delta_l}{\sigma_R}.
  \end{equation}
  
  \noindent The halo number density in Lagrangian coordinates is given by (see \eg Appendix C of Ref. \cite{Karagiannis2018}) 
  
  \begin{equation}\label{eq:delta_hL}
   \delta_h^L(M|M_1,V_0)=\frac{n_h(M|M_1,V_0,z)}{n_h(M,z)}-1.
  \end{equation}
  where $n_h(M|M_1,V_0)$ is the number of subhalos of mass $M$ with an initial volume $V_0$, corresponding to the small wavelength peaks ready to collapse on top of the long mode, above some mass $M_1$ defined by the ``background'' (\ie long wavelength) mode and $n_h(M,z)$ is the mean number of halos above mass $M$ (\ie the halo mass function). Taylor expanding \eref{eq:delta_hL} and comparing it with the local-in-matter bias expansion, we can identify the Lagrangian bias coefficients as
  
   \begin{align}\label{eq:PBS_arg}
   b_N^L(M,z)&=\frac{1}{n_h(M,z)}\frac{\partial^Nn_h(M,z)}{\partial\delta_l^N}\bigg|_{\delta_l=0}=\frac{(-\nu)^N}{\delta_c^Nf(\nu,z)}\frac{d^Nf(\nu,z)}{d\nu^N}\,.
  \end{align}
  
  This is a general result for any universal mass function, therefore we can use it to derive the halo bias parameters for the fitting mass function of Ref.~\cite{Tinker2008}, given by
  
  \begin{equation}\label{eq:Tinker_MF}
      f(\nu,z)=\alpha\left[1+(\beta\nu)^{-2\phi}\right]\nu^{2\eta}e^{-\gamma\nu^2/2},
  \end{equation}
  where the parameters have the following redshift dependence
  
  \begin{align}
      \beta(z)=\beta_0(1+z)^0.2, \\
      \phi=\phi_0(1+z)^{-0.08}, \\
      \eta=\eta_0(1+z)^0.27, \\
      \gamma=\gamma_0(1+z)^{-0.01},
  \end{align}
  where the zero in the subscript denotes the values of the fitting parameters for $z=0$ and can be found in Table 4 of Ref. \cite{Tinker2010}, together with amplitude $\alpha$ values.
  
  The first four local-in-matter Lagrangian halo bias parameters are

  \begin{align}
  b_1^L&=\frac{2 \varphi }{\delta_c\left[(\beta  \nu )^{2 \varphi }+1\right]}+\frac{\gamma  \nu
   ^2-2 \eta -1}{\delta_c}, \label{eq:b1h} \\
  b_2^L&=\frac{2 \varphi  \left(2 \gamma  \nu ^2-4 \eta +2 \varphi
   -1\right)}{\delta_c^2\left[(\beta  \nu )^{2 \varphi }+1\right]}+\frac{\gamma ^2 \nu ^4-4
   \gamma  \eta  \nu ^2-3 \gamma  \nu ^2+4 \eta ^2+2 \eta}{\delta_c^2}, \label{eq:b2h}
    \\
  b_3^L&=\frac{2 \varphi  \left(6 \varphi  \left(\gamma  \nu ^2-2 \eta
   \right)+3 \left(\gamma  \nu ^2-2 \eta \right)^2-6 \gamma  \nu
   ^2+4 \varphi ^2-1\right)}{\delta_c^3\left[(\beta  \nu )^{2 \varphi }+1\right]} \nonumber \\
   &+\frac{\gamma^3 \nu ^6-6 \gamma ^2 \eta  \nu ^4-6 \gamma ^2 \nu ^4+12 \gamma 
   \eta ^2 \nu ^2+12 \gamma  \eta  \nu ^2+3 \gamma  \nu ^2-8 \eta
   ^3+2 \eta}{\delta_c^3}, \label{eq:b3h}  \\
   b_4^L&=\frac{4 \varphi}{\delta_c^4\left[(\beta  \nu )^{2 \varphi}+1\right]}  \big[2 \gamma ^3 \nu ^6-2 \eta  \left(6
   \gamma ^2 \nu ^4+6 \gamma  \nu ^2 (2 \varphi -1)+8 \varphi ^2+6
   \varphi -1\right)+3 \gamma ^2 \nu ^4 (2 \varphi -3) \nonumber \\
   &+12 \eta ^2\left(2 \gamma  \nu ^2+2 \varphi +1\right)+\gamma  \nu ^2
   \left(8 \varphi ^2-6 \varphi +1\right)-16 \eta ^3+4 \varphi ^2
   (\varphi +1)-\varphi -1\big] \nonumber \\
   &+\frac{\gamma ^4 \nu ^8-8 \gamma ^3 \eta  \nu ^6-10 \gamma ^3 \nu^6+24 \gamma ^2 \eta ^2 \nu ^4+36 \gamma ^2 \eta  \nu ^4+15\gamma ^2 \nu ^4}{\delta_c^4} \nonumber \\
   &+\frac{-32 \gamma  \eta ^3 \nu ^2-24 \gamma  \eta ^2
   \nu ^2-4 \gamma  \eta  \nu ^2+16 \eta ^4-16 \eta ^3-4 \eta ^2+4
   \eta}{\delta_c^4}. \label{eq:b4h}
 \end{align}
 From the mapping between Eulerian and Lagrangian bias in the spherical collapse approximation, we get \footnote{See \eg Appendix C of Ref. \cite{Karagiannis2018} for more details.}

  \begin{figure*}
\centering
\resizebox{0.8\linewidth}{!}{\includegraphics{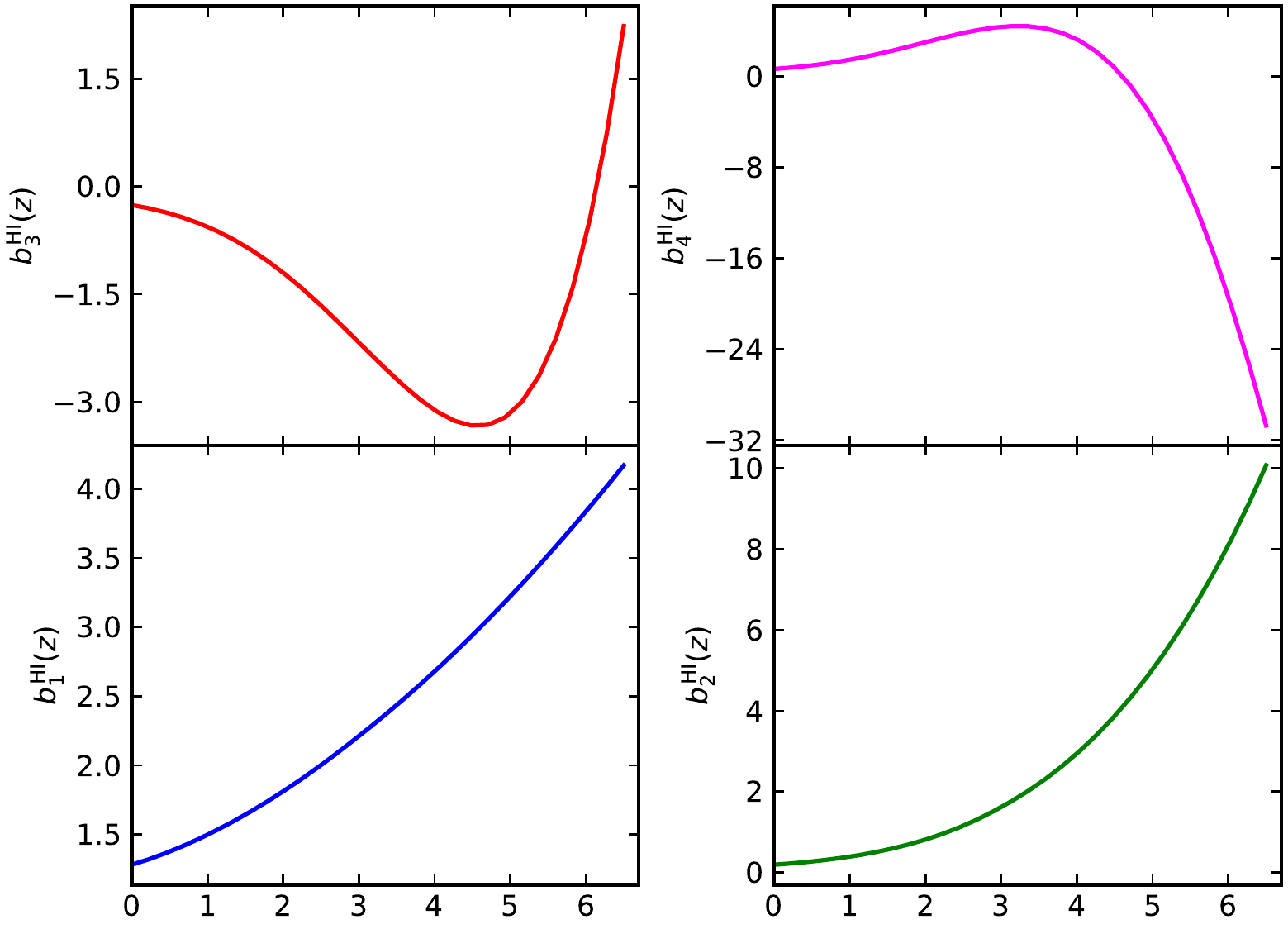}}
    \caption{The first four local-in-matter Eulerian bias parameters of the HI galaxies [\eref{eq:bHI}] as a function of redshift. The derivation uses the HOD model of Ref. \cite{Castorina2016}, which is briefly described in \sref{sec:HI_bias}. } \label{fig:bias_params}

\end{figure*}
  
  \begin{align}
   b_1^E&=1+b_1^L, \label{eq:b1hE} \\
   b_2^E&=b_2^L+2(\alpha_1+\alpha_2)b_1^L, \label{eq:b2hE} \\
   b_3^E&=6(\alpha_2 + \alpha_3)b_1^L+ 3(1 + 2\alpha_2)b_2^L +b_3^L, \label{eq:b3hE} \\
   b_4^E&=24(\alpha_3 + \alpha_4)b_1^L+ 12(\alpha_2^2 + 2(\alpha_2 + \alpha_3))b_2^L +4(1 + 3\alpha_2)b_3^L +b_4^L. \label{eq:b4hE}
  \end{align}
  where $\alpha_1=1$, $\alpha_2=-17/21$, $\alpha_3=2815/3969$ and $\alpha_4=-590725/916839$.
  
  The resulting Eulerian HI bias parameters, after using the methodology and HOD model described in \sref{sec:HI_bias}, are plotted in \fref{fig:bias_params}.

\section{Baseline distribution}
\label{app:baselines}

Accurate modeling of the effective distribution of baselines is non-trivial for transit telescopes. The sky rotation that takes objects over the sky has the effect that the same piece of sky is measured by the same baseline with different projection and the $m$-mode analysis (cite) demonstrates that some information that is not sampled directly can be recovered through time variation. In particular, for experiments like HIRAX, different patches of the sky will be observed with different pointing altitudes and PUMA will also likely to have at least one degree of freedom per dish. We skim over these details and use smoothed version of physical distribution of baseline lengths instead. 

 \begin{figure*}
\centering
\resizebox{0.75\linewidth}{!}{\includegraphics{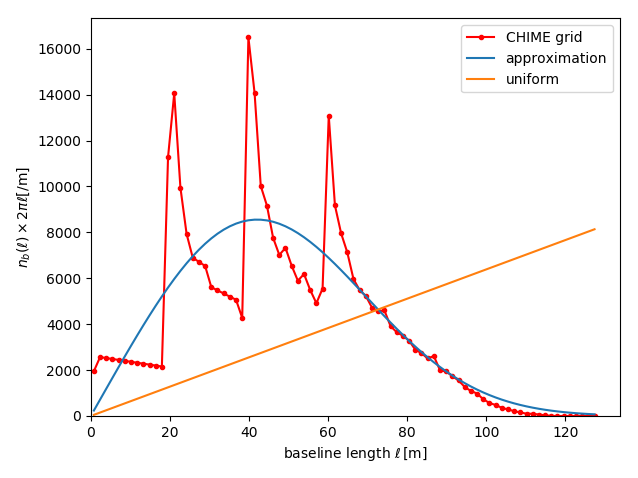}}
    \caption{A comparison of the CHIME baseline distribution with the approximating formula of \eref{eq:base_CHIME} and the uniform distribution, which assumes a constant sampling in the u-v plane (see \eg Ref. \cite{Bull2015}). } \label{fig:baselineplot}

\end{figure*}

   The baseline distribution for the PUMA and HIRAX surveys is given by Eq. D4 of \cite{StageII2018}, given by the following fitting formula
 \begin{equation}\label{eq:base_SII_XIR}
   n_{b}(l) = n_0\frac{a+b (l/L)}{1+c (l/L)^d} e^{-(l/L)^e},
 \end{equation}
 where $n_0=(N_s/D_{\rm dish})^2$ and $L=N_sD_{\rm dish}$ with
 $n_b(u) = n_{b}(l=u\lambda) \lambda^2$, while $N_s$ is the number of antennas in the side of the square array (\ie $N_s=256$ for PUMA and $N_s=32$ for HIRAX). This formula has been calibrated so that
 $\int n_b(u) d^2u = N_s^2/2$. The fitting parameters for a square closed-packing array, considered in the case of HIRAX, are
 $a=0.4847,\;b=-0.3300,\;c=1.3157,\;d=1.5974,\;e=6.8390$. For PUMA, as discussed in the main text, we consider a hexagonal closed-packing array in a compact cycle, where the fitting parameters are now $a=0.5698,\;b=-0.5274,\;c=0.8358,\;d=1.6635,\;e=7.3177$. For both cases, the fitted parameters can be found in Appendix D of Ref. \cite{StageII2018}, as well as a detailed discussion.

  For CHIME the baseline distribution is given by the following fitting formula
\begin{equation}\label{eq:base_CHIME}
n_b(l) = A\exp\left[-(l/B)^C\right],
\end{equation}
where the parameters are, $A=48.5511,\;B=60.693,\;C=2.4797$. A comparison between the actual baseline length distibution for CHIME and the fitting formula is shown in \fref{fig:baselineplot}.

\section{Redshift space kernels and the Finger-of-God dumping term} \label{app:RSD_kernels_FOG}

  The effect of redshift space distortions (RSD) \cite{Kaiser1987, Hamilton1998} can be treated perturbatively \cite{Verde1998,Scoccimarro1999}, generalising the kernel formalism of SPT in order to include the redshift distortions and the bias terms [Eqs. (\ref{eq:deltaG}) and (\ref{eq:deltaNG})]. The general non-Gaussian redshift kernels up to second order, while neglecting $\order{f_{\rm NL}^2}$ terms, are given by (see e.g. Refs.~\cite{Baldauf2011,Tellarini2016,Karagiannis2018}):
  
  \begin{align}
   &Z_1(\bk_i)=b_1+f\mu_i^2+\frac{b_{\Psi}k_i^{\alpha}}{M(k_i,z)}, \label{eq:Z1}\\
   &Z_2(\bk_i,\bk_j)=b_1F_2(\bk_i,\bk_j)+f\mu_{ij}^2G_2(\bk_i,\bk_j)+\frac{b_2}{2} +b_{s^2}S_2(\bk_i,\bk_j) \nonumber \\ 
   &+\frac{f\mu_{ij}k_{ij}}{2}\left[\frac{\mu_i}{k_i}Z_1(\bk_j)+\frac{\mu_j}{k_j}Z_1(\bk_i)\right] +\frac{1}{2}\left(\frac{(b_{\Psi\delta}-b_{\Psi}N_2(\bk_j,\bk_i))k_i^{\alpha}}{M(k_i,z)}+\frac{(b_{\Psi\delta}-b_{\Psi}N_2(\bk_i,\bk_j))k_j^{\alpha}}{M(k_j,z)}\right) \label{eq:Z2},
  \end{align}
  where $f$ is the linear growth rate, $\mu_i=\bk_i\cdot\hat{z}/k_i$ is the cosine of the angle between the wavevector $\bk_i$ and the line-of-sight $\hat{z}$, $\mu_{ij}=(\mu_ik_i+\mu_jk_j)/k_{ij}$ and $k_{ij}^2=(\bk_i+\bk_j)^2$. The kernels $F_2(\bk_i,\bk_j)$ and $G_2(\bk_i,\bk_j)$ are the second order symmetric kernels of SPT (see Ref.~\cite{Bernardeau2002} for a review), while $S_2(\bk_1,\bk_2) = (\bk_1\cdot\bk_2)^2/(k_1^2 k_2^2)-1/3$ and $N_2(\bk_1,\bk_2) = (\bk_1\cdot\bk_2)(k_1^2)$. The $S_2$ kernel arises from the Fourier transform of the tidal field scalar $s^2$ \citep{McDonald2009,Baldauf2012}, while $N_2$, in the presence of PNG, originates from the coupling of the displacement field between the Eulerian and Lagrangian frames to the primordial gravitational potential \cite{Giannantonio2010,Baldauf2011}. For the different PNG shapes considered here, parameter $\alpha$ get the following values: $\alpha=2$ for the equilateral shape, $\alpha=1$ for the orthogonal configuration, while the usual local case can be retrieved by $\alpha=0$ \cite{Dalal2008,Slosar2008,Giannantonio2010}. Note that in the expressions above, the redshift dependence has been suppressed for clarity.

  The Finger-of-God (FOG) is taken into account here, where the damping effect of the clustering power is described phenomenologically  by \cite{Peacock1994,Ballinger1996} 
  
    \begin{align}
  &D_\text{FOG}^P(\bk)=e^{-(k\mu\sigma_P)^2} \label{eq:DfogP}, \\
  &D_\text{FOG}^B(\bk_1,\bk_2,\bk_3)=e^{-(k_1^2\mu_1^2+k_2^2\mu_2^2+k_3^2\mu_3^2)\sigma_B^2}. \label{eq:DfogB} 
  \end{align}
  Here we consider the fiducial values for $\sigma_P = \sigma_B = \sigma_{\upsilon}(z)$, where $\sigma_{\upsilon}$ is the usual linear, one dimensional velocity dispersion. Note that, due to the high precision of the redshift measurements in 21 cm IM surveys considered here, the redshift error is assumed to be zero, \ie perfect knowledge of redshift.

\end{document}